\documentclass[twoside,twocolumn,9pt]{article}
\usepackage{extsizes}
\usepackage[super,sort&compress,comma]{natbib} 
\usepackage[version=3]{mhchem}
\usepackage[left=1.5cm, right=1.5cm, top=1.785cm, bottom=2.0cm]{geometry}
\usepackage{balance}
\usepackage{times,mathptmx}
\usepackage{sectsty}
\usepackage{graphicx} 
\usepackage{lastpage}
\usepackage[format=plain,justification=justified,singlelinecheck=false,font={stretch=1.125,small,sf},labelfont=bf,labelsep=space]{caption}
\usepackage{float}
\usepackage{fancyhdr}
\usepackage{fnpos}
\usepackage[english]{babel}
\usepackage{array}
\usepackage{droidsans}
\usepackage{charter}
\usepackage[T1]{fontenc}
\usepackage[usenames,dvipsnames]{xcolor}
\usepackage{setspace}
\usepackage[compact]{titlesec}
\usepackage{tikz}
\usetikzlibrary{positioning}

\usepackage{xfrac}
\usepackage{amsmath}
\usepackage[mathscr]{eucal}

\newcommand{\RN}[1]{%
	\textup{\uppercase\expandafter{\romannumeral#1}}%
}
\usepackage[caption=true]{subfig}
\usepackage{epstopdf}
\definecolor{cream}{RGB}{222,217,201}

\begin{document}

\pagestyle{fancy}
\thispagestyle{plain}
\fancypagestyle{plain}{

\fancyhead[C]{\includegraphics[width=18.5cm]{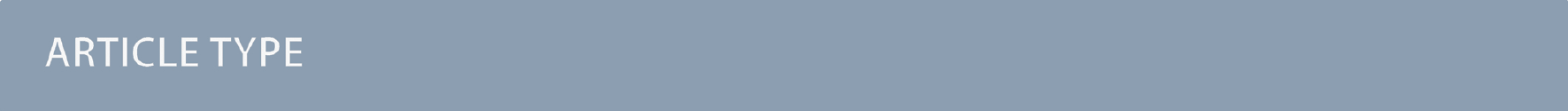}}
\fancyhead[L]{\hspace{0cm}\vspace{1.5cm}\includegraphics[height=30pt]{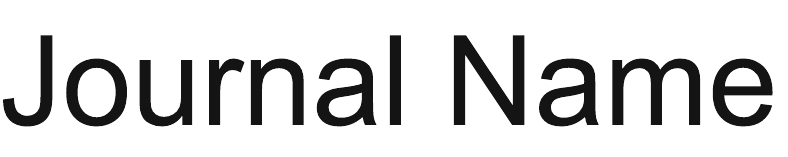}}
\fancyhead[R]{\hspace{0cm}\vspace{1.7cm}\includegraphics[height=55pt]{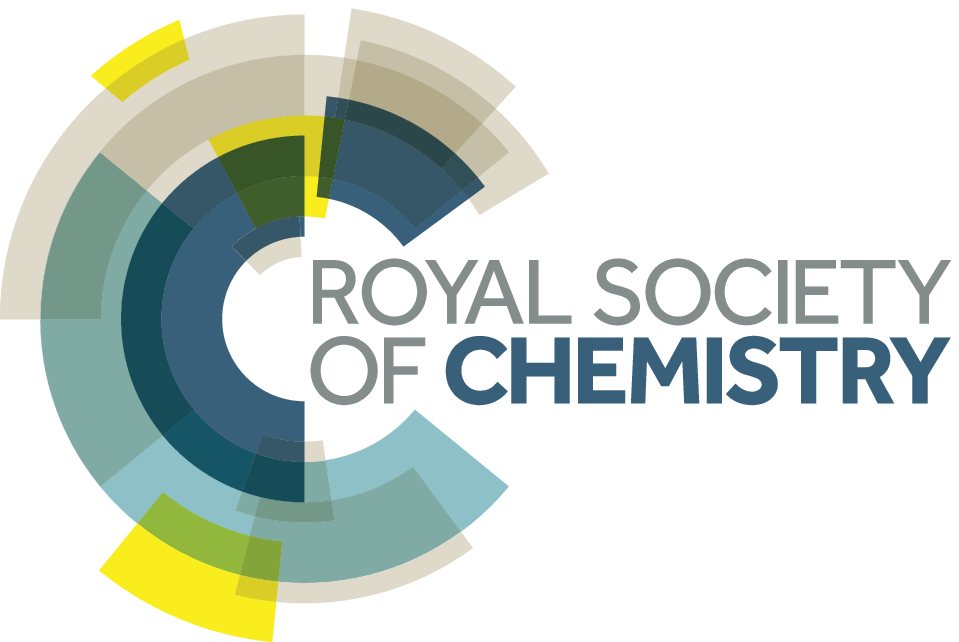}}
\renewcommand{\headrulewidth}{0pt}
}

\makeFNbottom
\makeatletter
\renewcommand\LARGE{\@setfontsize\LARGE{15pt}{17}}
\renewcommand\Large{\@setfontsize\Large{12pt}{14}}
\renewcommand\large{\@setfontsize\large{10pt}{12}}
\renewcommand\footnotesize{\@setfontsize\footnotesize{7pt}{10}}
\makeatother

\renewcommand{\thefootnote}{\fnsymbol{footnote}}
\renewcommand\footnoterule{\vspace*{1pt}%
\color{cream}\hrule width 3.5in height 0.4pt \color{black}\vspace*{5pt}} 
\setcounter{secnumdepth}{5}

\makeatletter 
\renewcommand\@biblabel[1]{#1}            
\renewcommand\@makefntext[1]%
{\noindent\makebox[0pt][r]{\@thefnmark\,}#1}
\makeatother 
\renewcommand{\figurename}{\small{Fig.}~}
\sectionfont{\sffamily\Large}
\subsectionfont{\normalsize}
\subsubsectionfont{\bf}
\setstretch{1.125} 
\setlength{\skip\footins}{0.8cm}
\setlength{\footnotesep}{0.25cm}
\setlength{\jot}{10pt}
\titlespacing*{\section}{0pt}{4pt}{4pt}
\titlespacing*{\subsection}{0pt}{15pt}{1pt}

\fancyfoot{}
\fancyfoot[LO,RE]{\vspace{-7.1pt}\includegraphics[height=9pt]{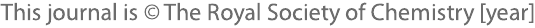}}
\fancyfoot[CO]{\vspace{-7.1pt}\hspace{13.2cm}\includegraphics{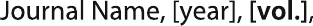}}
\fancyfoot[CE]{\vspace{-7.2pt}\hspace{-14.2cm}\includegraphics{head_foot/RF}}
\fancyfoot[RO]{\footnotesize{\sffamily{1--\pageref{LastPage} ~\textbar  \hspace{2pt}\thepage}}}
\fancyfoot[LE]{\footnotesize{\sffamily{\thepage~\textbar\hspace{3.45cm} 1--\pageref{LastPage}}}}
\fancyhead{}
\renewcommand{\headrulewidth}{0pt} 
\renewcommand{\footrulewidth}{0pt}
\setlength{\arrayrulewidth}{1pt}
\setlength{\columnsep}{6.5mm}
\setlength\bibsep{1pt}

\makeatletter 
\newlength{\figrulesep} 
\setlength{\figrulesep}{0.5\textfloatsep} 

\newcommand{\topfigrule}{\vspace*{-1pt}%
\noindent{\color{cream}\rule[-\figrulesep]{\columnwidth}{1.5pt}} }

\newcommand{\botfigrule}{\vspace*{-2pt}%
\noindent{\color{cream}\rule[\figrulesep]{\columnwidth}{1.5pt}} }

\newcommand{\dblfigrule}{\vspace*{-1pt}%
\noindent{\color{cream}\rule[-\figrulesep]{\textwidth}{1.5pt}} }

\makeatother

\twocolumn[
  \begin{@twocolumnfalse}
\vspace{3cm}
\sffamily
\begin{tabular}{m{4.5cm} p{13.5cm} }

\includegraphics{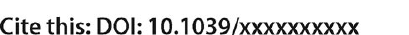} & \noindent\LARGE{\textbf{Phase separation dynamics of polydisperse colloids:\newline 
a mean-field lattice-gas theory}} \\
\vspace{0.3cm} & \vspace{0.3cm} \\

 & \noindent\large{Pablo de Castro$^{\ast}$ and Peter Sollich} \\

\includegraphics{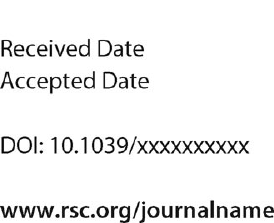} & \noindent\normalsize{New insights into phase separation in colloidal suspensions are provided via a new dynamical theory based on the Polydisperse Lattice-Gas model. The model gives a simplified description of polydisperse colloids, incorporating a hard-core repulsion combined with polydispersity in the strength of the attraction between neighbouring particles. Our mean-field equations describe the local concentration evolution for each of an arbitrary number of species, and for an arbitrary overall composition of the system. We focus on the predictions for the dynamics of colloidal gas-liquid phase separation after a quench into the coexistence region. The critical point and the relevant spinodal curves are determined analytically, with the latter depending only on three moments of the overall composition. The results for the early-time spinodal dynamics show qualitative changes as one crosses a `quenched' spinodal that excludes fractionation and so allows only density fluctuations at fixed composition. This effect occurs for dense systems, in agreement with a conjecture by Warren that, at high density, fractionation should be generically slow because it requires inter-diffusion of particles. We verify this conclusion by showing that the observed qualitative changes disappear when direct particle-particle swaps are allowed in the dynamics. Finally, the rich behaviour beyond the spinodal regime is examined, where we find that the evaporation of gas bubbles with strongly fractionated interfaces causes long-lived composition heterogeneities in the liquid phase;
we introduce a two-dimensional density histogram method that allows such effects to be easily visualized for an arbitrary number of particle species.} 


\\

\\

\end{tabular}

 \end{@twocolumnfalse} \vspace{0.6cm}

  ]

\renewcommand*\rmdefault{bch}\normalfont\upshape
\rmfamily
\section*{}
\vspace{-1cm}


\footnotetext{\textit{Disordered Systems Group, Department of Mathematics, King's College London, London, United Kingdom. E-mail: pablo.decastro@kcl.ac.uk}}
\footnotetext{\dag~Electronic Supplementary Information (ESI) available: Animation shows evolution of phase-separation dynamics snapshots for a binary mixture (left) and their two-dimensional density histograms (right). Parameters: $p^A=p^B=0.435$, $T=0.5$, $d=0.25$, $w_0=1$, $w_s=0$, $L=150$, from $t=0$ to $t=4497.5$. See DOI: 10.1039/b000000x/ \label{esi}}


\section{Introduction}
Traditionally, thermodynamics and statistical mechanics deal with systems composed of identical particles, also called \textit{monodisperse} systems. However, nature is often more complex: many soft matter systems, such as biological and industrial fluids (in the form of colloidal dispersions, liquid crystals, polymer solutions, etc.), are \textit{polydisperse} in the sense that their constituent particles exhibit variation in terms of one (or several) attribute(s).\cite{sollich2011polydispersity,auer2001suppression,liddle2011polydispersity,poon2002physics,mohanty2014effective,tan2014visualizing,zaccarelli2015polydispersity,chrysikopoulos2014effect,van2009experimental,stuart1980polydispersity,belli2011polydispersity,sluckin1989polydispersity,sollich2005nematic,rogovsic1996polydispersity,vieville2011polydispersity} This \textit{polydisperse attribute}, denoted here by $\sigma$, can be particle size, shape, charge, molecular weight, chemical nature, etc.\cite{evans1998universal} In the thermodynamic limit, these systems are usually regarded as having effectively an infinite number of \textit{components} (although the theory that will be presented here applies to systems with an \textit{arbitrary} number of components).\cite{0953-8984-14-3-201,bartlett1998soft} Everyday and not-so-everyday examples include blood, paint, milk, clay, photonic crystals, shampoo, viruses, globular proteins, pharmaceuticals, and even sewage, among many others.\cite{berger1991polydisperse,jose1996maya,anema2008effect,johnson1955particle,lavrinenko2009influence,li2003direct,fraden1995phase,gazzillo2011effects,vera1996colloidal,park2009spectroscopic} Therefore, understanding the impact of polydispersity on the \textbf{phase behaviour} of many-body systems is of fundamental, commercial, and practical interest.\cite{szymanska2015stability} For instance, knowing under what conditions (and how) a multicomponent fluid will demix may be essential in determining the shelf life of a product.\cite{rayner2015engineering,lietor2009role} On the fundamental end of the range, it is believed that (symmetry-breaking) phase transitions have occurred in the early universe via nucleation of vacuum-field bubbles, thus generating the different fundamental forces of nature; despite having a different sort of fluid dynamics and not being in the scope of the present work, it is interesting that the early-universe hot plasma can also be seen as a multicomponent fluid.\cite{ahonen1998transport,langlois2008perturbations} Our focus here is on colloidal fluids, for which the fluid particles are influenced by the thermal agitation from the solvent in which they are suspended, and quantum effects can be neglected. (See ref.\ \citenum{williamson2013kinetics} for an introduction to polydisperse colloidal systems and their phase behaviour.)

The state of a polydisperse system (or any of its phases) is described by a density distribution $\rho(\sigma)$, defined such that $\rho(\sigma)d\sigma$ is the number density of particles with $\sigma$-values in the range $\left[\sigma,\sigma+d\sigma\right]$. One can also consider a multicomponent system, composed of a finite number $M$ of particle species (each one labelled by its own value of $\sigma$). Each component has concentration $p^{\alpha}$, with $\alpha=1,2, \dots ,M$. (Note that the $\alpha$ here is a superscript, not an exponent.) The continuous approach can be thought of as the limit $M\rightarrow\infty$, leading to $p^{\alpha}\rightarrow \rho(\sigma_{\alpha})d\sigma$, where $\sigma_{\alpha}$ is the value of $\sigma$ associated with species $\alpha$. In either case, the system has an overall \textit{composition}, which specifies the ratios of the densities of different species.
The simplest example is the one of a binary system, i.e.\ $M=2$: its composition could then be specified by the ratio $p^A/p^B$, for instance. We will often specialize to this binary case for ease of explanation.



\begin{figure}[h]
\centering
\includegraphics[clip,width=\columnwidth]{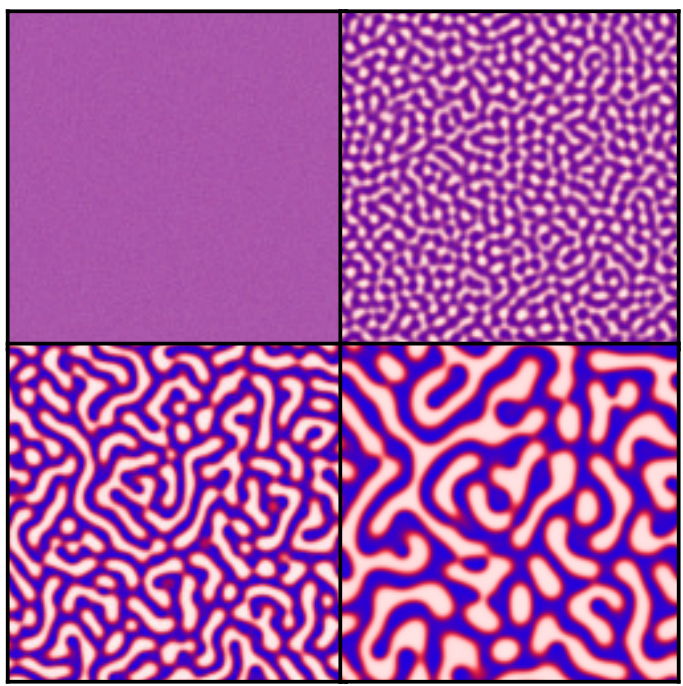}%
\caption{Phase separation snapshots (from top left to bottom right) for a binary mixture (generated with our equations)\textemdash see text for details, particularly Section \ref{binumerics}.}
\label{illustrative}
\end{figure}

Similarly to monodisperse fluids, polydisperse fluids can phase-separate into regions with higher and lower concentrations of particles. The typical experiment is to decrease the temperature of a homogeneous system to a value within the coexistence region, 
setting off a dynamical evolution towards separation into two (or more) equilibrium phases. (See Fig.\ \ref{illustrative} and the animation provided in the ESI.\dag) In polydisperse phase separation, however, these phases will not just differ in density but generally also in composition. This process is called \textit{fractionation} (or \textit{partitioning}) \cite{0953-8984-14-3-201} because it implies that particles of different species distribute themselves unevenly into the new phases.
Fractionation is responsible for much of the complexity in the phase behaviour of polydisperse systems;
\cite{0953-8984-14-3-201,PhysRevE.77.011501,evans,Fairhurst2004} its effect on phase-separation kinetics will be our main focus.

\begin{figure}[h]
\centering
\includegraphics[clip,width=0.9\columnwidth]{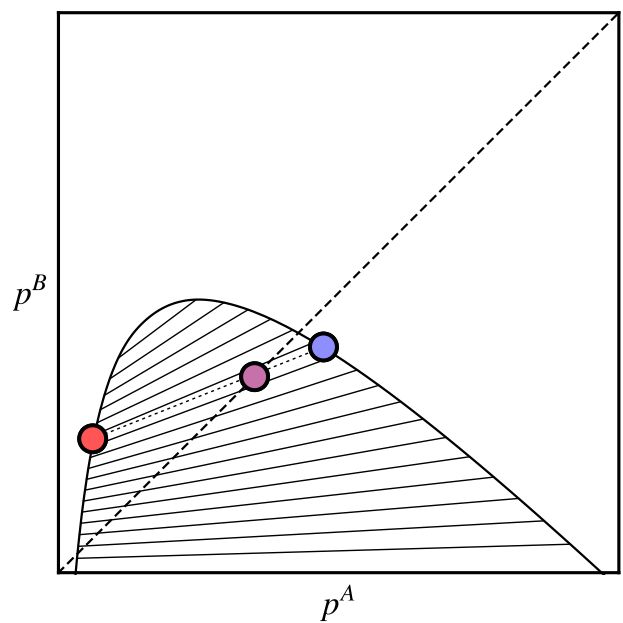}%
\caption{Sketch of a binary fluid phase diagram at fixed temperature. The purple (centre) point represents the parent phase. The points to the right and to the left (high- and low-density phases, respectively) are the two daughter phases. A \textit{tieline} connects the daughter phases, defining the \textit{equilibrium fractionation direction}; it passes through the parent phase. A number of other example tielines are shown. The end points of all tielines, taken together, form the boundary of the coexistence region.}%
\label{fracbi}
\end{figure}

Let us explain fractionation in more detail using the $M=2$ example. Fig.\ \ref{fracbi} shows a schematic example of a phase diagram of a binary fluid at a fixed temperature. The purple (middle) point indicates the species densities $(p^A, p^B)$ of the \textit{parent phase}, i.e.\ of the initial homogeneous system. Any other phase of the same composition would lie on the dashed line through the parent and the origin. This is called the  \textit{dilution line} because in a colloidal fluid such phases can be prepared by adding or removing solvent. Phase separation can occur into pairs of coexisting phases -- identified from the requirements of equal pressure and species chemical potentials -- that are shown as end points of tielines. 
Due to particle conservation, the overall system composition must remain unchanged during phase separation, so the actual \textit{daughter phases} generated must lie on the tieline passing through the parent.
In the generic case these daughter phases both lie off the dilution line so have a composition that is different from the parent (and from each other). This is the phenomenon of fractionation.


In the case of general $M$ similar considerations apply. 
Here the parent is specified by a density distribution $\rho^{(0)}(\sigma)$. This can be decomposed as
$\rho^{(0)}(\sigma)=\rho^{(0)}_{0}f^{(0)}(\sigma)$, where $\rho^{(0)}_{0}=N/V$ is the overall particle number density and $f^{(0)}(\sigma)$, the normalized parent shape function, specifies the composition. As $\rho^{(0)}_{0}$ is varied, $\rho^{(0)}(\sigma)$ traces out the dilution line in density distribution space. To obtain phase diagrams
one needs to project from this $M$-dimensional space. Often only the density of coexisting phases is recorded, to recover the polydisperse analogue of a monodisperse density--temperature phase diagram (see Fig.~\ref{phasediagram}).

Fractionation causes the equilibrium phase behaviour of polydisperse systems to be much richer than that of their monodisperse counterparts.\cite{0953-8984-14-3-201}  
To see why, note that for large  $M$, the classical tangent plane procedure used to find phase coexistence from a given free energy becomes awkward both conceptually and computationally.
Also, the number of coexisting phases allowed by Gibbs' phase rule grows with $M$, and becomes arbitrary in the limit $M\to\infty$.
These challenges have been tackled using a number of theoretical methods, uncovering many qualitatively novel effects of polydispersity on phase behaviour. \cite{0953-8984-14-3-201,fasolo2004fractionation,sollich2010crystalline,bartlett1999reentrant} For instance, Evans has developed a perturbative approach for narrow distributions of the polydisperse attribute.\cite{evans1998universal,evans,evans2001diffusive,crystalgrowth,sollich2008weakly} This predicts, for example, that
the difference in mean particle size between the two daughter phases is proportional to the variance of the parent distribution.
It can also predict specific trends, e.g.\ that in the fluid-solid coexistence of size-disperse colloidal hard spheres, the solid (crystalline) phase contains on average larger particles than the fluid.\cite{evans}
Going beyond perturbative approaches, the full equilibrium phase diagram for polydisperse hard spheres has been found via accurate free energy expressions for the fluid and solid phases, showing very different equilibrium properties already for moderate particle size spreads.\cite{fasolo2003equilibrium} The physical intuition behind this is the fact that particles of different sizes cannot easily fit into a regular crystal lattice. 
These results were derived using a moment projection method.\cite{0953-8984-14-3-201,sollich1998projected,warren1998combinatorial,momentfreeenergies}
This exploits a mapping of the full density distribution space onto \textit{moment densities}, defined as $\rho_n \equiv \int \sigma^n \rho(\sigma) d\sigma$, where one integrates over all possible values of $\sigma$. Note that the zeroth moment density $\rho_0$ is just the overall density $N/V$. Another important parameter, the standard deviation of the polydisperse attribute $\sigma$ normalized by its mean, can be expressed in terms of moment densities as $[(\rho_2 \rho_0/\rho_1^2) - 1]^{1/2}$. It is often referred to as \textit{degree of polydispersity} or simply \textit{polydispersity}.

The present work focuses on the \textit{dynamics} towards equilibrium. Warren proposed that the phase-separation kinetics in polydisperse systems should proceed in two stages.\cite{A809828J} To understand this two-stage scenario, he starts by considering an initially homogeneous system (suddenly placed within a two-phase region). The classical picture predicts two types of phase-separation kinetics. If placed not too far from the phase boundary encompassing the two-phase region in the phase diagram, the system would phase-separate via \textit{nucleation and growth}. However, if placed further away from the boundary phase separation would proceed via \textit{spinodal decomposition}. Lying entirely below the coexistence curve, the classical spinodal curve separates the two types of kinetics. In the case of phase separation via spinodal decomposition, arbitrarily small fluctuations in the homogeneous system decrease the free energy, and thus grow exponentially before eventually being limited by nonlinear effects. Warren highlights the fact that, although the spinodal curve is a convention that does not rigorously survive in a correct statistical mechanical treatment, it is a rough guide to the transition region between the two types of kinetic behaviour. He proceeds by asking what the effect of polydispersity is on this classical picture. Assuming that in dense systems, fractionation is potentially a very slow process, he suggests that two stages of kinetics might be expected.
In a first stage, the system might phase-separate by relaxing the overall density to a phase equilibrium dictated by `quenching' (i.e.\ `fixing') the polydisperse distribution in any phase to the one in the parent phase. Over longer timescales, the system might then gradually redistribute particles between coexisting phases to reflect the possibility of further lowering the free energy by fractionating. Therefore, a second stage would begin when the system started to significantly fractionate. Referring to the polydisperse density distribution as the size distribution, Warren says: `During this slow process, the particle number density would be able to adjust itself continually to follow the current stage of the size partitioning process. The argument for this [two-stage] scenario is based on the observation that, \textit{in a dense system}, the overall density can be much more easily relaxed by collective particle motion (collective diffusion), than can the size distribution be adjusted by individual particle rearrangements (self diffusion)' (p.\ 2199). Experimental measurements of self- and collective diffusion constants support this picture.\cite{pusey1986phase}

Warren's two-stage scenario is certainly physically reasonable: density fluctuations can be created by moving groups of particles `in sync' to form regions of higher density. Fractionation, on the other hand, requires particles from different species to `push past each other' in opposite directions.
The scenario also makes an interesting connection to moment densities: the
zeroth moment $\rho_0$, which gives the total number density of all species, should relax much more rapidly at a local level than the higher moments, whose equilibration requires interdiffusion of different particle species. This suggests that moment densities can remain
useful in understanding the kinetics of phase separation, and our results in this paper provide some support for this.

Existing theoretical approaches to polydisperse phase-separation dynamics have focussed primarily on polymeric systems. 
One could make some progress by binning the range of $\sigma$, which reduces the problem to the dynamics of a finite mixture. This approach of course becomes numerically challenging as the number of bins grows. 
Clarke has shown that the method is nonetheless useful for investigating the early-time phase-separation dynamics of polydisperse polymers.\cite{0953-8984-14-3-201,Clarke2001} Again in the context of polymeric materials, Pagonabarraga and Cates developed an analysis of the dynamics based on time evolution equations for polymer densities driven by chemical potential gradients. 
The form of these equations had been proposed phenomenologically in Clarke's work, but the approach of ref.~\citenum{ignacio} allowed the associated mobility coefficients to be derived explicitly. Apart from the case of length polydispersity, where the results were more subtle, Pagonabarraga and Cates studied the case of chemical polydispersity, where different polymer chains have different monomer compositions and hence different interaction strengths. For this scenario the coupled dynamical equations could be fully solved in certain cases. Pagonabarraga and Cates\cite{ignacio} also studied the mode spectrum of the various density fluctuations in a system undergoing spinodal decomposition; from this analysis, they concluded that depending on where in the phase diagram the system is placed the kinetics will proceed in accordance with Warren's two-stage scenario.

Compared to the polymeric case, there is a lack of theoretical work designed to model fractionation effects in the phase-separation dynamics of spherical \textit{colloids} (but see ref.~\citenum{evans2001diffusive} below). This paper is designed to fill this gap.
The approach we use to investigate the phase-separation dynamics of mixtures is the theory described in refs.\ \citenum{PhysRevLett.78.4970} and \citenum{Plapp99} by Plapp and Gouyet. These studies were, however, concerned with \textit{binary metallic alloys}. In this context they addressed rather different questions from ours, based on assumptions about the dynamics and the particle interactions that are quite distinct from the colloidal case.
Nonetheless, our development of the mean-field dynamical equations has close similarities with the methods of refs.~\citenum{PhysRevLett.78.4970} and \citenum{Plapp99}.


There are a few other theoretical investigations of polydisperse colloidal dynamics in the literature, where simulations are implemented and some aspects of fractionation investigated.\cite{measuringfractionation,Square-well,crystalgrowth} As a result, the dynamics of phase separation in polydisperse colloids remains a challenging (and mostly unsolved) problem. 
One of the difficulties is that the kinetics of phase separation could be so slow as to make the actual equilibrium phase compositions unobservable in experiments.\cite{0953-8984-14-3-201} It has been argued
that this is the case for polydisperse hard-sphere crystals: once particles join to a crystal nucleus growing from the hard-sphere fluid, they essentially no longer diffuse on experimental timescales.\cite{evans2001diffusive} 
The size distribution of particles in the crystal will thus `freeze in', and will be determined by the mechanism of crystal growth rather than the conditions of thermodynamic equilibrium. Although recent advances have been made,\cite{liddle2014polydispersity} such non-equilibrium effects on the experimentally observed phase behaviour of colloidal systems are definitely not fully understood.

In this paper we present a mean-field theory for the Polydisperse Lattice-Gas (PLG) model, which has been proposed as a simple description of polydisperse colloids.\cite{PhysRevE.77.011501} In Section \ref{PLG} we present the model and its mean-field phase diagram. In Section \ref{kinetic} we endow the model with an appropriate dynamics and derive the mean-field evolution equations for this. Section \ref{linear} gives an early-time regime analysis of our equations. Then, in Section \ref{numerics}, we go beyond the spinodal regime and study the full late-time dynamics of our model. Section \ref{conclusion} summarizes our results and outlines some directions for future research. 

\section{PLG mean-field phase diagram}
\label{PLG}
\subsection{PLG model}
Hereafter we use the PLG model to investigate the phase-separation dynamics in polydisperse colloidal systems.\cite{PhysRevE.77.011501} The model is described by the Hamiltonian
\begin{equation}
H = - \sum\limits_{\langle i,j\rangle}^{}\sum\limits_{\alpha, \beta}^{}\sigma_{\alpha}\sigma_{\beta}n_{i}^{\alpha}n_{j}^{\beta}
\label{plg}
\end{equation}
where $i$ runs over the sites of a periodic lattice $i =1, \dots ,L^{D}$, assumed simple cubic and $D$-dimensional in this work, with lattice spacing $a=1$, unless otherwise stated; the sum runs over all pairs $\langle i,j\rangle$ of nearest-neighbour sites; $\sigma_{\alpha}$ is the value of the polydisperse attribute associated with particle species $\alpha$, which controls the strength of interparticle interactions; it is a positive number for attractive interactions as considered here. We consider a mixture with $M$ species, with the summations over $\alpha$ and $\beta$ therefore running from $1$ to $M$. The (occupation) variable $n_{i}^{\alpha}$ simply counts the number of particles of species $\alpha$ at site $i$, for which a hard-core constraint is imposed:
\begin{equation}
\sum\limits_{\gamma=0}^{M} n_{i}^{\gamma}=1\quad \forall i
\label{hardcore}
\end{equation}
where $n_i^{0}$ refers to \textit{vacancy}, i.e.\ $n_i^{0}=1$ indicates the presence of a vacancy at site $i$, or, equivalently, a solvent particle. Note that the solvent particles are `passive'\cite{ignacio} in this framework, in such a way that any non-hydrodynamic effect caused by them has already been effectively included in the interaction between the particles as described by the model Hamiltonian.
We also neglect the fact that 
colloidal particles may interact with one another via hydrodynamic interactions mediated by the solvent.
 We therefore use the term vacancy in the following, rather than solvent particle.
Observe that $n_i^{0}$ can be expressed in terms of the other $n_i^{\alpha}$, from eqn (\ref{hardcore}). In summary, each lattice site may be either vacant or occupied by a single colloidal particle of polydisperse attribute $\sigma_\alpha$. The instantaneous density distribution follows as $p^\alpha= L^{-D}\sum_{i}^{}n_{i}^{\alpha}$. We will use the letter $\gamma$ as the species index for summations running from $0$ to $M$, while for summations running from $1$ to $M$, we use $\alpha$ (or $\beta$), unless otherwise specified.

Note that in the Hamiltonian (\ref{plg}) the interaction strength between any two neighbouring particles is assumed to be $\sigma_\alpha \sigma_\beta$, i.e.\ the product of their polydisperse attributes, though to preserve generality we will often write this in the form $\epsilon_{\alpha\beta}=\sigma_{\alpha}\sigma_{\beta}$. Thus the role of polydispersity in this model is to engender a spread of possible interaction strengths between particles, a situation which contrasts with the single interaction strength characterizing the simple Ising lattice-gas model. As observed in ref.\ \citenum{PhysRevE.77.011501}, this allows the PLG model to capture the essential qualitative features that distinguish polydisperse fluids from their monodisperse counterparts. Nonetheless one has to bear in mind that, in real colloids, polydispersity often occurs in the {\em size} of the particles. This will have additional consequences, e.g. on the local packing of particles in dense regions, that a lattice model cannot account for. In principle one could extend the approach by allowing particles to occupy several contiguous lattice sites, thus explicitly representing their size. This is not a trivial extension as particle moves would then correspond to simultaneous changes of potentially many $n_i^\alpha$, but may be an interesting avenue for future work.

\subsection{Mean-field phase diagram}
We next present the mean-field phase diagram for the PLG model. This will serve as a useful reference point for our later discussion of the phase-separation dynamics. We first explain how to obtain the relevant curves of the diagram, starting from the spinodals and then moving on to the cloud (`binodal') and shadow curves.

\subsubsection{Free energy and spinodals}

Phase separation via spinodal decomposition occurs when the system is placed in a region of the phase diagram that is \textit{unstable} to fluctuations;  fluctuations of any size will then lead to phase separation.
This contrasts with the case of nucleation and growth, where finite fluctuations -- corresponding to a nucleus of a critical size -- are required for the system to escape from a  \textit{metastable} state. 

For the phase diagram of monodisperse systems, the spinodal curve (i.e.\ the curve below which spinodal decomposition occurs) can be calculated by joining up the inflection points of the free energy curve at each temperature. For the spinodal curve of a  polydisperse system, one likewise needs to identify points where the free energy develops negative curvature. 

In order to obtain an expression for the Helmholtz (inhomogeneous) free energy of the polydisperse lattice gas,  $F=\left\langle H \right\rangle -TS$, we use a variational mean-field approximation.  This is obtained from the Gibbs\textendash Bogoliubov\textendash Feynman variational bound,\cite{kuzemsky2015variational,feynman1998statistical} using a variational approximation to the equilibrium distribution that is factorized over lattice sites. The latter is fully characterized by local densities $p_{i}^{\alpha}=\left\langle n_{i}^{\alpha} \right\rangle$. For the PLG Hamiltonian (\ref{plg}), this leads to
\begin{equation}
F=- \sum\limits_{\langle i,j\rangle}^{}\sum\limits_{\alpha, \beta}^{}\epsilon_{\alpha\beta} p_{i}^{\alpha}p_{j}^{\beta}+T\sum\limits_{i}^{}\sum\limits_{\gamma=0}^{M}p_{i}^{\gamma}\ln{p_{i}^{\gamma}}
\label{inhomogeneous}
\end{equation}
where the Boltzmann constant has been set equal to $1$. Recalling that the volume of the system is equal to the total number of sites ($V=L^D$) and applying eqn (\ref{inhomogeneous}) to a homogeneous configuration, the \textit{free energy density} $f=F/V$ can be written as
\begin{equation}
f(\left\lbrace p^{\alpha} \right\rbrace,T)=-\frac{z}{2}\sum\limits_{\alpha,\beta}^{}\epsilon_{\alpha\beta}p^{\alpha}p^{\beta}+T\sum\limits_{\gamma=0}^{M}p^{\gamma}\ln{p^{\gamma}}
\label{freeenergydens}
\end{equation}
where $z$ is the lattice coordination number [i.e.\ $z=4$ for the square lattice (in $D=2$)] and $p^{\alpha}$ is the overall density of particles of species $\alpha$. Here the hard-core constraint (\ref{hardcore}) means that in the second, entropic term, $p^0=1-\sum_\alpha p^\alpha$.
For a monodisperse system (i.e.\ $M=1$), the free energy density  reduces to a function $f=f(\rho,T)$, where $\rho=N/V\equiv p^1$ is the overall density of particles. Here and in the following we drop the 0-subscript from our previous notation $\rho_0$ where the meaning is clear from the context.
(Confusion with the density distribution 
$\rho(\sigma)$ should not arise as this is always written with its polydisperse attribute argument $\sigma$.) In this monodisperse case, the spinodal density at any temperature $T$ would be found by
from the equation $\frac{\partial^2f}{\partial\rho^2}=0$. In the more general, polydisperse, case one needs to consider the Hessian matrix $S$ with elements
\begin{equation}
S_{\alpha\beta}=\frac{\partial^2f}{\partial p^\alpha \partial p^\beta}
\end{equation}
and then solve the equation $\det(S)=0$ (or the equivalent spinodal criterion as written in ref.\ \citenum{momentfreeenergies}) to obtain the spinodal curve.

For polydisperse systems, one can, in the spirit of Warren's two-stage scenario, define two types of spinodals.\cite{A809828J} The first is the \textit{annealed} spinodal curve, which is precisely the one given by $\det(S)=0$. For the PLG model, this leads to a simple expression in terms of the moment densities:
\begin{equation}
	T=z\left(\rho_2-\rho_1^2\right).
	\label{annealedspinodal}
\end{equation}
The \textit{quenched} spinodal curve,  on the other hand, is defined as the spinodal curve that is calculated by assuming that the system is \textit{not} allowed to fractionate. In other words, one treats the composition as fixed, and the overall density $\rho$ as the only variable. To calculate this, note that the free energy given by eqn (\ref{freeenergydens}) is a function of all densities $p^\alpha$. At fixed composition these are proportional to the overall density $\rho$, e.g.\ for $M=2$ one would have $p^A=p^B=\rho/2$ if the dilution line $p^A/p^B=1$ is considered. Inserting these $\rho$-dependencies into the free energy gives the quenched free energy, as a function of $\rho$; call this $f_Q(\rho,T)$, for example. To find the quenched spinodal one then only has to solve $\frac{\partial^2f_Q}{\partial\rho^2}=0$, which in our case yields
\begin{equation}
 T=z \left(\frac{\rho_1}{\rho}\right)^2\left(\rho-\rho^2\right).
\end{equation}
One can easily check that this coincides with eqn (\ref{annealedspinodal}) for monodisperse systems, where $\rho_2/\rho = (\rho_1/\rho)^2$.

The above distinction between two types of spinodal will prove useful later because, if Warren's two-stage scenario holds,
one would
expect that the dynamics of phase separation would proceed, at least initially, as if the system `did not know' it could further lower its free energy by fractionating. Therefore, one could expect that the early-stage dynamics of a system operating in the two-stage scenario should proceed in accordance with the quenched spinodal instead of the annealed one.

For completeness we note that the critical point lies on the (annealed) spinodal curve and is identified by an additional condition. This can be  obtained using the methods of ref.\ \citenum{momentfreeenergies} and reads $2\rho_1^3-3\rho_1\rho_2+\rho_3=0$.

\subsubsection{Cloud and shadow curves}

Because of fractionation, the conventional vapour-liquid binodal (or coexistence curve) of a monodisperse system splits into a \textit{cloud curve}, marking the onset of (polydisperse) phase coexistence, and a \textit{shadow curve}, giving the density of the incipient phase. The critical point appears at the intersection of these curves, rather than at the maximum of either. \cite{0953-8984-14-3-201} This splitting is seen in experiments on polydisperse fluids (see e.g.\ ref.\ \citenum{MACP:MACP021951003}). Similarly to the spinodal curve, we can define annealed and quenched cloud curves. These will be distinct because the onset of phase coexistence will be delayed to lower temperature if the system is not allowed to lower its free energy by fractionation.

Our numerical data for cloud and shadow curves are determined by solving the equations for two-phase  (bulk) phase equilibrium. (Three-phase coexistence can also occur in the PLG, but at lower temperatures than we consider here.) In the binary case ($M=2$), the phase equlibrium conditions at given $T$ are equality of the chemical potentials $\mu^A$, $\mu^B$ and the pressure between the two phases 
$(p^{A}_{\thinspace\thinspace\!_\RN{1}}, p^{B}_{\thinspace\thinspace\!_\RN{1}})$ and $(p^{A}_{\thinspace\!_\RN{2}}, p^{B}_{\thinspace\!_\RN{2}})$.
Here the chemical potentials $\mu^A$ and $\mu^B$ are given by the partial derivatives of the free energy density $f(p^A,p^B,T)$ with respect to $p^A$ and $p^B$, respectively, and the pressure is $P=-f+\mu^A p^A +\mu^B p^B$.
For $M>2$, the generalization is straightforward, giving $M+1$ coexistence equations to solve. \big(Alternatively, one could use the moment free energy method, since our free energy density (\ref{freeenergydens}) is \textit{truncatable}.\cite{momentfreeenergies}\big) The cloud temperature for a given parent density is found by lowering $T$ and checking when phase coexistence with the parent as one of the coexisting phases is first found. The shadow curve identifies the density of the second coexisting phase, which at this point is present only in an infinitesimal fraction of the system volume. In a monodisperse system, cloud and shadow curve coincide and are then identical to the conventional binodal curve.

\subsubsection{Phase diagram}

\begin{figure}[t]
	\centering	
	\subfloat[]{%
		\includegraphics[width=\columnwidth]{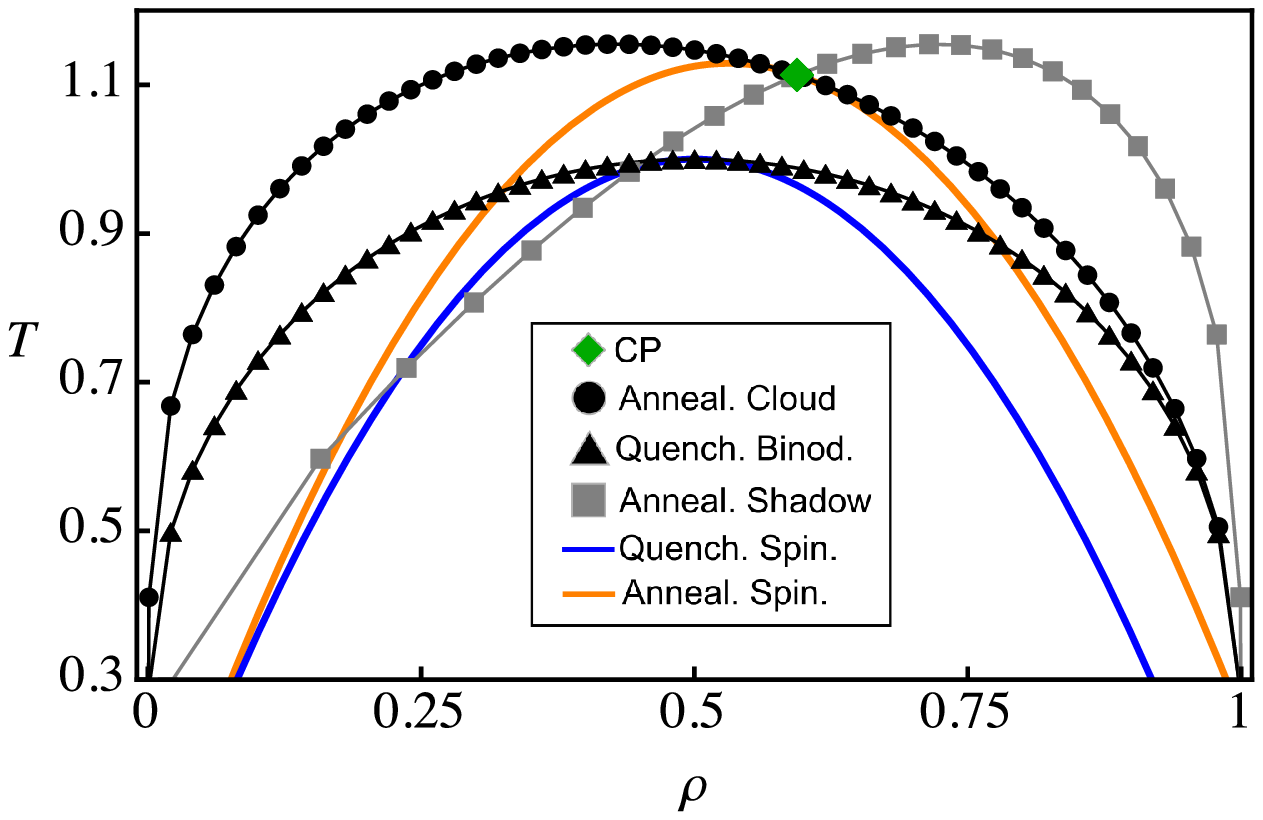}%
		\label{phasediagram}
	}
	
	\centering	
	\subfloat[]{%
		\includegraphics[width=0.9\columnwidth]{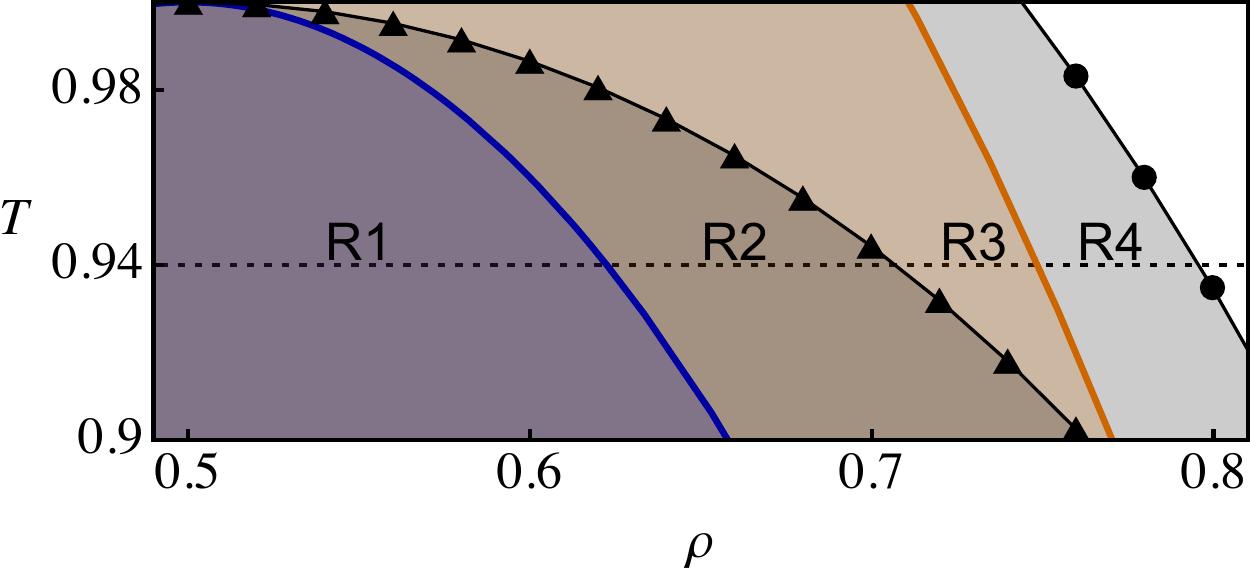}%
		\label{regions}
	}
	\caption{(a) Phase diagram of a binary PLG mixture for $d=0.25$ and dilution line $p^A=p^B$. These choices allow all the relevant qualitative features to be seen clearly.
		(b) Starting at $\rho=0.5$ and moving towards higher $\rho$ along a reference horizontal dotted line (at $T=0.94$), one visits four distinct phase coexistence regions with respect to the positions of the cloud/binodal and spinodal curves. According to Table \ref{thetab}, each of these regions corresponds to a distinct dynamical behaviour.
		}
\end{figure}

Fig.\ \ref{phasediagram} shows our results for the phase diagram of a binary mixture. We chose the dilution line $p^A=p^B$ and $\sigma$-values given by $\sigma_{A}=1+d$ and $\sigma_{B}=1-d$, where $d$ is a number between $0$ and $1$. (Note that the greater the $d$, the greater the polydispersity in this bidisperse case.) The plot shows the annealed cloud and shadow curves, with the critical point at their intersection; the quenched binodal curve; and the annealed and quenched spinodal curves.
Qualitatively these curves look as one would expect them to on general grounds.\cite{0953-8984-14-3-201} 

For our discussion of phase-separation dynamics most relevant are the cloud curve as it signals the onset of phase coexistence, and the spinodal curve. The annealed and quenched versions of each of these two curves divide the phase diagram into distinct regions where Warren's hypothesis predicts different sequences of phase-separation dynamics. Inside the quenched spinodal, the system should initially phase-separate by spinodal decomposition in density only; between the quenched spinodal and quenched binodal the first stage should be nucleation and growth of density fluctuations. The second stage would then involve fractionation, again by spinodal decomposition or nucleation and growth depending on the position relative to the annealed spinodal.
 
As an example, Fig.\ \ref{regions} shows a zoomed-in portion of the phase diagram. Starting at $\rho=0.5$ and moving towards higher densities at constant temperature
$T=0.94$, one visits four distinct regions with respect to the annealed and quenched cloud and spinodal curves. Labelling these by R1, R2, R3, R4 with increasing density gives the predictions in Table~\ref{thetab},
where we
have used the term `size' generically to refer to the polydisperse attribute $\sigma$.
In this paper we develop a mean-field theory that excludes fluctuations, hence does not capture nucleation and growth dynamics. 
Hence we will focus on the distinction between different spinodal regions, i.e.\ region R1 versus R2/R3. Note that we will use the term `early-time dynamics' throughout to refer to the onset of phase-separation dynamics, irrespective of the relevant time scale. In particular, in region R2/R3 the early-time dynamics should be stage 2 spinodal decomposition, which happens on a slow time scale because it involves fractionation. 

\begin{table}[h]
\centering
	\begin{tabular}{| c | c | c |}
		\hline
		\textbf{Region} & \textbf{Stage 1} & \textbf{Stage 2} \\
		\hline
		R1 & SD in density & SD of `sizes' \\
		\hline
		R2 & NG of density fluct.\ &
		SD of `sizes' \\
		\hline
		R3 & ---  & SD of `sizes' \\
		\hline
		R4 & --- & NG of `size' fluct.\\
		\hline
	\end{tabular}
\caption{Abbreviations: SD = Spinodal Decomposition; NG = Nucleation and Growth; fluct.\ = fluctuations.}
\label{thetab}
\end{table}

\section{Kinetic PLG model}
\label{kinetic}
In this section we present the  kinetics we assume for the PLG model with an arbitrary number of species $M$, and derive our mean-field dynamical equations. 

We model the dynamics as resulting from jumps of particles to nearest-neighbour sites. This can be described in the general form of a master equation
\begin{equation}
\frac{\partial P(C,t)}{\partial t}=\sum\limits_{C'}^{}\left[ W(C',C)P(C',t)-W(C,C')P(C,t)\right]
\end{equation}
where $P(C,t)$ is the probability of finding the system in a configuration $C=\lbrace n_i^{1}, n_i^{2}, \dots , n_i^{M}; i=1, \dots , L^D \rbrace$ at time $t$, and $W (C,C')$ is the transition rate from a configuration $C$ to another one $C'$. One can then define a time-dependent average for any observable ${Q}(C)$
as
\begin{equation}
\langle Q \rangle (t)=\sum_{C}^{}Q(C)P(C,t)
\end{equation}
In particular we 
define a time-dependent local density for each species as $p_i^{\alpha}(t)=\langle n_i^{\alpha} \rangle (t)$. From the master equation it can be shown that each species follows a conservation law of the form
\begin{equation}
\frac{dp_i^{\alpha}}{dt}=-\sum\limits_{j\in \partial i}^{}J_{ij}^{\alpha}
\label{conservationlaw}
\end{equation}
where we have introduced the notation $j\in\partial i$, meaning that the summation has to be performed over all nearest-neighbour sites $j$ of
site $i$; the current of $\alpha$-particles across the link $i\rightarrow j$ is given by
\begin{equation}
J_{i j}^{\alpha}=\sum\limits_{\gamma=0}^{M}\Bigl\langle n_i^{\alpha} n_j^{\gamma}	w_{i j}^{\alpha \gamma} - n_j^{\alpha}n_i^{\gamma}	w_{ji}^{\alpha \gamma}\Bigr\rangle
\label{current}
\end{equation}
where $w_{i j}^{\alpha\gamma}$ is the \textit{jump rate} for an $\alpha$-particle at site $i$ to exchange positions with a $\gamma$-particle at site $j$ (or with a vacancy, if $\gamma=0$). Observe that, effectively, we are considering a kinetic model equipped with two elementary processes: (i) the jump of a particle from an occupied lattice site to an empty one, and also (ii) the direct interchange between two particles from arbitrary species. Throughout this process, the configuration of the system changes, but the overall number of particles of each species remains constant, which is the physical origin of the conservation law (\ref{conservationlaw}).  It is worth noting that the physical elementary process are the jumps to empty sites, i.e.\ particle-vacancy exchanges. However, for the subsequent analysis it is useful to include also direct particle-particle swaps, so that we develop the theory initially for generic $w^{\alpha\gamma}_{ij}$. 

Note from that the prefactors of $w^{\alpha\gamma}_{ij}$ in eqn (\ref{current}) ensure the target site is occupied by a $\gamma$-particle (or empty, if $\gamma=0$) and the start site is occupied by an $\alpha$-particle. This is similar to the theory described by Plapp and Gouyet in ref.\ \citenum{PhysRevLett.78.4970}, but there they did not consider direct interchange of particles, and the number of species $M$ was set equal to $2$ (as previously mentioned); moreover, they considered Arrhenius law jump rates, which was argued to be more suitable in their context of phase separation in binary metallic alloys. Here, we use Glauber-like jump rates:
\begin{equation}
w_{i j}^{\alpha\gamma}=w^{\alpha\gamma}\left[\frac{1}{1+\exp{\left(\Delta H_{ij}^{\alpha\gamma}/T\right)}}\right]
\end{equation}
where $\Delta H_{ij}^{\alpha\gamma} = H(C')-H(C)$ is the energy difference associated with the jump, i.e.\ it is the energy difference between the configuration before the jump ($C$), in which there is an $\alpha$-particle at site $i$ and a $\gamma$-particle at site $j$, and after the exchange ($C'$); note that $C$ and $C'$ are identical except for the exchange between $\alpha$ and $\gamma$. The prefactor $w^{\alpha\gamma}$ is an `attempt rate'. This gives the actual jump rate $w_{i j}^{\alpha\gamma}$ when the energy change $\Delta H_{ij}^{\alpha\gamma}$ is large and negative, while it is reduced exponentially for large positive energy changes.


Let us write the energy of the system for the configuration before the jump, isolating only
the contributions involving the two particles $\alpha$ and $\gamma$ that are swapping:
\begin{equation}
H(C) = - \epsilon_{\alpha\gamma} - \sum\limits_{\beta}^{}\sum\limits_{k\in \partial i\backslash j}^{}\epsilon_{\alpha\beta}n_{k}^{\beta}-\sum\limits_{\beta}^{}\sum\limits_{l\in \partial j \backslash i}^{}\epsilon_{\gamma\beta}n_{l}^{\beta}
+\dots
\label{before}
\end{equation}
Here the first term on the right-hand side is the contribution only from the interaction between the two particles that are swapping. The second term is the contribution from the interactions between the $\alpha$-particle at site $i$ and all of its neighbours except the one at site $j$. The third term similarly accounts for the interactions of the $\gamma$-particle.
Similarly, we have that
\begin{equation}
H(C') = - \epsilon_{\alpha\gamma} - \sum\limits_{\beta}^{}\sum\limits_{k\in \partial i\backslash j}^{}\epsilon_{\gamma\beta}n_{k}^{\beta}-\sum\limits_{\beta}^{}\sum\limits_{l\in \partial j\backslash i}^{}\epsilon_{\alpha\beta}n_{l}^{\beta}+ \dots
\label{after}
\end{equation}
is the energy of the system after the exchange.

The energy change  
is then
\begin{equation}
\begin{aligned}
\Delta H_{ij}^{\alpha\gamma} ={}
& \sum\limits_{\beta}^{}\left(\sum\limits_{k\in \partial i\backslash j}^{}\epsilon_{\alpha\beta}n_{k}^{\beta}-\sum\limits_{l\in \partial j\backslash i}^{}\epsilon_{\alpha\beta}n_{l}^{\beta}\right)\\ 
&-\sum\limits_{\beta}^{}\left(\sum\limits_{k\in \partial i\backslash j}^{}\epsilon_{\gamma\beta}n_{k}^{\beta}-\sum\limits_{l\in \partial j\backslash i}^{}\epsilon_{\gamma\beta}n_{l}^{\beta}\right).
\end{aligned}
\label{12}
\end{equation}
We now invoke a mean-field approximation 
to evaluate the currents (\ref{current}). In this we neglect fluctuations 
of $\Delta H_{ij}^{\alpha\gamma}$, i.e.\ we replace this quantity by its average. This can be justified by thinking about a high-dimensional limit, where the number of nearest neighbour sites $z$ of any lattice site is large enough for the local `fields' appearing in $\Delta H_{ij}^{\alpha\gamma}$ to average out fluctuations. In the same spirit we also drop the restrictions on the sums in $\Delta H_{ij}^{\alpha\gamma}$, which only changes the local fields by a relative amount $1/z$. It then only remains to perform the average of the kinetic prefactors, which with a mean-field decoupling becomes $\langle n_i^{\alpha} n_j^{\gamma}\rangle \approx p_i^\alpha p_j^\gamma$.
Inserting the resulting approximation for the currents into (\ref{conservationlaw}), we obtain the mean-field kinetic equations
\begin{equation}
\begin{aligned}
\frac{dp_i^{\alpha}}{dt}=-\sum\limits_{j\in\partial i}^{}\sum\limits_{\gamma=0}^{M}\left[ \frac{p_i^{\alpha} p_j^{\gamma}	w^{\alpha\gamma}}{1+\exp{\left(\bigl\langle\Delta H_{ij}^{\alpha\gamma}\bigr\rangle/T\right)}}- \frac{p_j^{\alpha}p_i^{\gamma}	w^{\alpha\gamma}}{1+\exp{\left(\bigl\langle\Delta H_{ji}^{\alpha\gamma}\bigr\rangle/T\right)}}\right]
\end{aligned}
\label{kineticequations}
\end{equation}
where -- within the mean-field approximation --
\begin{equation}
\begin{aligned}
\bigl\langle\Delta H_{ij}^{\alpha\gamma}\bigr\rangle ={}
& \sum\limits_{\beta}^{}\left(\sum\limits_{k\in \partial i}^{}\epsilon_{\alpha\beta}p_{k}^{\beta}-\sum\limits_{l\in \partial j}^{}\epsilon_{\alpha\beta}p_{l}^{\beta}\right)\\ 
&-\sum\limits_{\beta}^{}\left(\sum\limits_{k\in \partial i}^{}\epsilon_{\gamma\beta}p_{k}^{\beta}-\sum\limits_{l\in \partial j}^{}\epsilon_{\gamma\beta}p_{l}^{\beta}\right).
\end{aligned}
\label{deltaHMF}
\end{equation}
These kinetic equations are -- in spite of the rather different method of derivation -- consistent with the mean-field free energy (\ref{inhomogeneous}) in the sense that they always 
decrease it over time, $dF/dt\leq 0$. \big(This consistency is what requires the approximation step we have taken above, of dropping the restrictions on the sums defining $\Delta H_{ij}^{\alpha\gamma}$.\cite{Puri}\big)
This is as one would expect in a closed system, where there are no currents crossing boundaries. \cite{Plapp99}
We defer the derivation of the result $dF/dt$ 
to Appendix \ref{monoticity}, which generalizes similar derivations discussed elsewhere in the literature\cite{Plapp99} 
to include the case of direct particle-particle swapping.

From the fact that $dF/dt\leq 0$ it follows that the dynamics leads to a state which minimizes the mean-field free energy. This final state may be the ground state (global minimum) or a metastable state (local minimum). The monotonic decrease of the free energy also implies that the mean-field kinetic equations cannot describe nucleation events. Capturing these would require introducing fluctuations, e.g.\ by adding Langevin noise to our deterministic eqns (\ref{kineticequations}). (This is further discussed in Section \ref{conclusion}.)

\section{Early-time spinodal dynamics}
\label{linear}
In this section, we will present a linearized version of our theory: it describes the growth of small fluctuations around an initial homogeneous state (via spinodal decomposition), within the framework of our mean-field kinetic equations. It will be shown that the maximum spinodal growth rates can be expressed in terms of only three moments of the polydisperse distribution. More importantly, we will use the result for the spinodal growth rates to test Warren's two-stage hypothesis.

We begin by considering a homogeneous system of overall composition described by a list of densities:
$\left\lbrace p^{\alpha} \mid \alpha=1,\dots,M \right\rbrace$. The system is perturbed by small fluctuations of the densities:
\begin{equation}
p^{\alpha}_i=p^{\alpha}+\delta^{\alpha}_{i}
\end{equation}
where $\delta^{\alpha}_{i}\ll1$. As shown in Appendix \ref{growthratesappendix}, linearization of (\ref{kineticequations}) leads to the following equation:
\begin{equation}
\frac{d\delta_i^{\alpha}}{dt}=\sum\limits_{\gamma=0}^{}\mathscr{M}^{\alpha\gamma}\Delta_{\rm d}\Bigl(\mu_{i}^{\alpha}-\mu_{i}^{\gamma}\Bigr)
\label{LinearDMFT}
\end{equation}
where we define the \textit{homogeneous mobilities} as
\begin{equation}
\mathscr{M}^{\alpha\gamma}\equiv\frac{w^{\alpha\gamma}}{2T}p^{\alpha}p^{\gamma}
\label{mob}
\end{equation}
We have also introduced the local chemical potentials $\mu_i^\alpha=\partial F/\partial p_i^\alpha$. These are given explicitly by
\begin{equation}
\mu_i^\alpha=-\sum_{j\in \partial i}\sum_{\beta}\epsilon_{\alpha\beta}p^{\beta}_j+T\ln{\left( p^{\alpha}_i/p^{0}_i\right) }
\label{chempot}
\end{equation}
for $\alpha=1,\dots,M$, while $\mu_i^0=0$. These expressions are derived from the free energy expression (\ref{inhomogeneous}) with the explicit substitution $p_i^0 = 1-\sum_\alpha p_i^\alpha$. 
Finally in eqn (\ref{LinearDMFT}) we use the \textit{discrete Laplacian} $\Delta_{\rm d}$, which is defined by 
\begin{equation}
\Delta_{\rm d}g_i=\sum\limits_{j\in \partial i}^{}\left(g_{j}-g_i\right)
\end{equation}
 for any site-dependent quantity $g$.
 
Note that, so far, the theory applies for completely generic attempt rates  $w^{\alpha\gamma}$. Furthermore, if these attempt rates are set such that the right-hand side of eqn (\ref{LinearDMFT}) contains only the particle-vacancy term (i.e.\ no direct interchanges of particles are allowed), and one considers only the $M=2$ case, then eqn (\ref{LinearDMFT}) has the same form as eqn (27) in Plapp and Gouyet's work.\cite{Plapp99} The only difference is that their expression for the homogeneous mobility is different from (\ref{mob}), as Plapp and Gouyet used Arrhenius jump rates. Their choice reflects the physical assumption for alloys that the elementary particle moves have an energy barrier effectively equivalent to removing a particle from the system. For the colloidal case Glauber rates are rather more plausible.

To solve the linearized mean-field equations, one exploits that a homogeneous system is invariant under translation with respect to the lattice vectors. Solutions are therefore superpositions of time-dependent Fourier modes
\begin{equation}
\delta_{j}^{\alpha}=\delta p^{\alpha}\exp\left[i\mathbf{k}\cdot\mathbf{x}_j+\omega t\right]
\label{planewave}
\end{equation}
(as will be clear, the $i$ in the exponents refers to the imaginary unit $i\equiv\sqrt{-1}$). Here $\mathbf{k}$ is the fluctuation wave vector and $\mathbf{x}_j$ is the position vector in real space of lattice site $j$. Moreover, $\omega$ is the \textit{growth rate} of the mode and $\delta p^{\alpha}$ indicates the amplitude of the fluctuation associated with species $\alpha$. By inserting eqn (\ref{planewave}) into (\ref{LinearDMFT}) one finds an eigenvalue equation with eigenvalue $\omega$, with the $\delta p^{\alpha}$ being the components of the corresponding $M$-dimensional eigenvector. [See eqn (\ref{omegaalpha}) in Appendix \ref{growthratesappendix}.] Thus we have a stability spectrum with $M$ branches. A branch can be stable ($\omega$ is negative for all wave vectors) or unstable, with $\omega$ being positive in some range of $|\mathbf{k}|$, typically for small $|\mathbf{k}|$.
We will be interested in the \textit{maximum growth rate} $\omega_{\rm max}$ over all branches and wave vectors, which identifies the dominant growing fluctuation mode.
Outside of the spinodal region this maximum growth rate 
becomes zero because the system is stable to all small fluctuations there.

To be able to evaluate the maximum growth rate we need to make specific assumptions about the attempt rates $w^{\alpha\gamma}$. We will set $w^{\alpha0}=w_0$ and $w^{\alpha\beta}=w_s$, for any $\alpha\neq0$ and $\beta\neq0$, where $w_0$ and $w_s$ are constant attempt rates associated with particle-vacancy and particle-particle exchanges, respectively. (The `$0$' subscript is for vacancy, and the `$s$' is for \textit{swapping}.) In principle, one could imagine that $w^{\alpha\beta}$ might not be the same for all pairs $\alpha$ and $\beta$, or that it depends on the temperature. However, our simple choice for the values of $w^{\alpha\gamma}$ is enough to distinguish between particle-vacancy and particle-particle kinetic mechanisms. Moreover, we will see later that the dependence on temperature would be irrelevant for our purposes. Also, Plapp and Gouyet say in ref.\ \citenum{Plapp99} that, in their case with particle-vacancy dynamics only (where $M=2$ and Arrhenius rates are used), numerical results indicate that qualitative phase-separation behaviour is unaffected by the attempt rate ratio $w^{A0}/w^{B0}$ as long as it is not too far from unity. 


For the above choice of $w^{\alpha\gamma}$, we show in Appendix \ref{growthratesappendix}, that the (largest branch of) growth rates can be expressed as
\begin{equation}
\begin{aligned}
\omega ={}
&\frac{A(\mathbf{k})}{4T}
\left\lbrace T\left[\left(2-\rho\right)w_0+w_s\rho\right]-\left(A(\mathbf{k})+z\right)\biggl[w_0\left(1-\rho\right)\rho_2\right.\\
&\left.+w_s\left(\rho\rho_2-\rho_1^2\right)\biggr]\right\rbrace-\frac{A(\mathbf{k})}{4T}\biggl[\left(T\left(w_0-w_s\right)\rho\right)^2\\
&+\left(A(\mathbf{k})+z\right)^2\left(w_0\left(1-\rho\right)\rho_2+w_s\left(\rho\rho_2-\rho_1^2\right)\right)^2\biggr.\\
&+2T\left(w_0-w_s\right)\left(A(\mathbf{k})+z\right)\left(w_0\left(1-\rho\right)\left(\rho\rho_2-2\rho_1^2\right)\right.\\
&\biggl.\left.+w_s\rho\left(\rho\rho_2-\rho_1^2\right)\right)\biggr]^{1/2}
\end{aligned}
\label{growthrates}
\end{equation}
where $A(\mathbf{k})$ is essentially the Fourier transform of the 
Laplacian. Equation (\ref{growthrates}) is valid for any spatial dimension given the appropriate expression for $A(\mathbf{k})$, with e.g.\ for a two-dimensional lattice $A(\mathbf{k})=-4\sin^2\left(k_{x}a/2\right)-4\sin^2\left(k_{y}a/2\right)$. The moment densities in the above expressions are, in the discrete representation, $\rho_{n}=\sum\limits_{\alpha}^{}\sigma_\alpha^{n}  p^{\alpha}$.
 
\begin{figure}[t]
\hspace{-0.0cm}
\subfloat[]{%
  \includegraphics[clip,width=0.5\columnwidth]{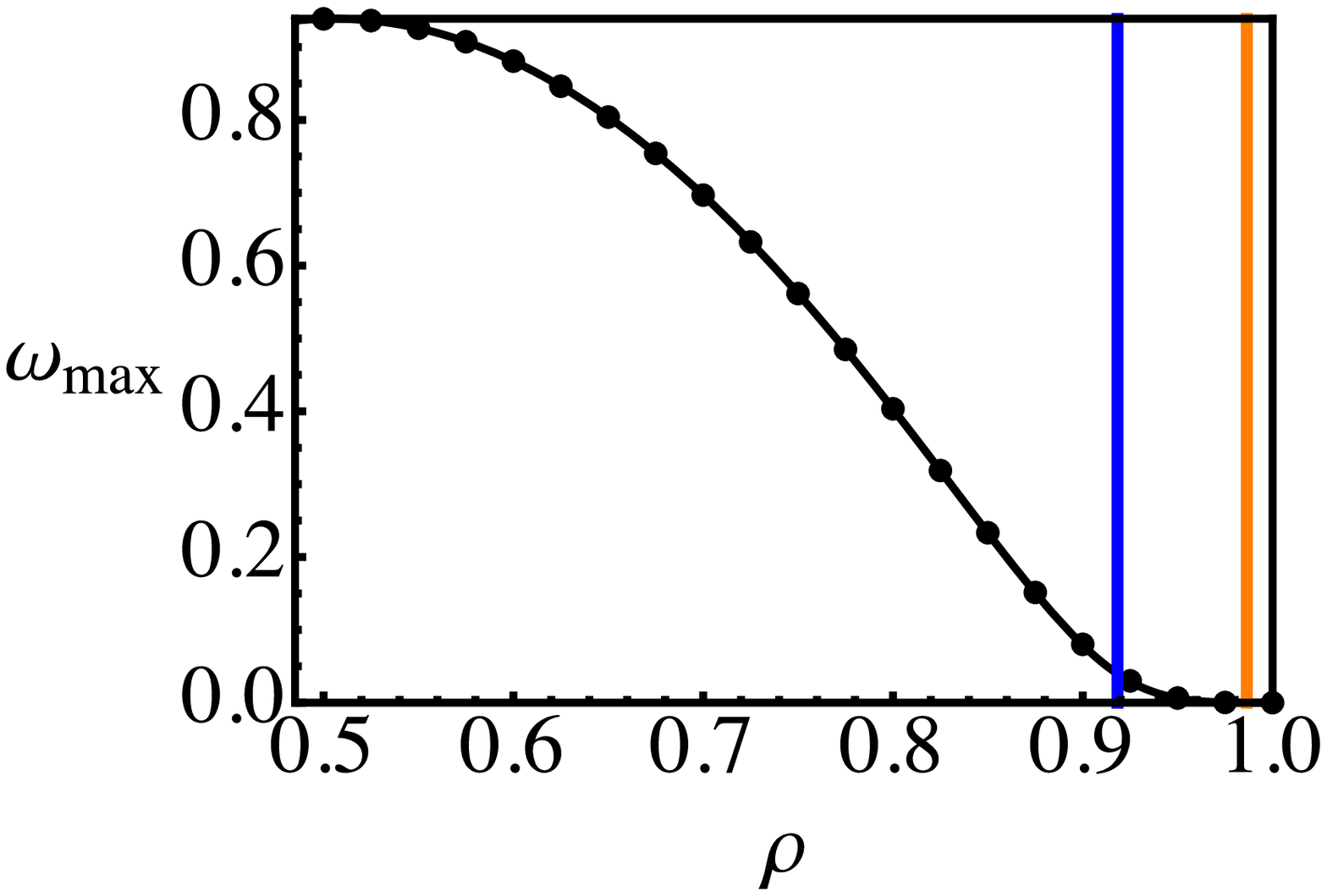}%
\label{wmaxa}}
\hspace{0.1cm}
\subfloat[]{\hspace{-0.25cm}\includegraphics[clip,width=0.5\columnwidth]{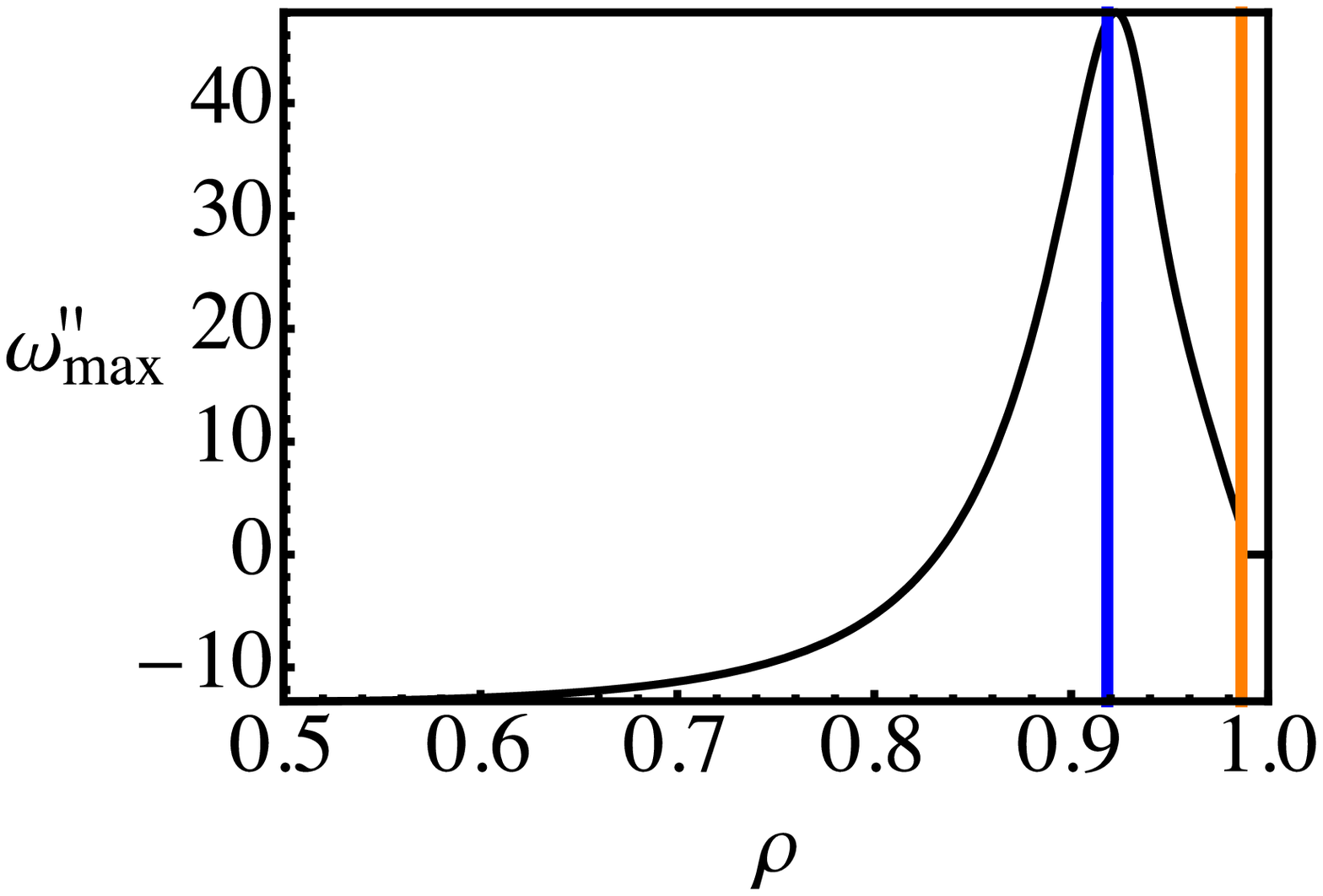}%
\label{wmaxb}}
\caption{(a) $\omega_{\rm max}$ as function of $\rho$ for $T=0.3$, $g=15$ (standard deviation: 25\% of the mean), and $w_s=0$. The vertical lines indicate the (upper) quenched and annealed spinodal densities, from left to right. For comparison, we show the points obtained by numerically maximising $\omega(\mathbf{k})$ and the curve obtained analytically from small-$|\mathbf{k}|$ expansion; these agree very well. (b) Second density derivative of $\omega_{\rm max}$ as function of $\rho$. The vertical lines are the same as in (a). Note that $\omega_{\rm max}$ almost has a kink at the quenched spinodal density, as indicated by the maximum in the second derivative.}
\label{wmax}
\end{figure}

\begin{figure}[h!]
\subfloat[]{%
  \includegraphics[clip,width=0.5\columnwidth]{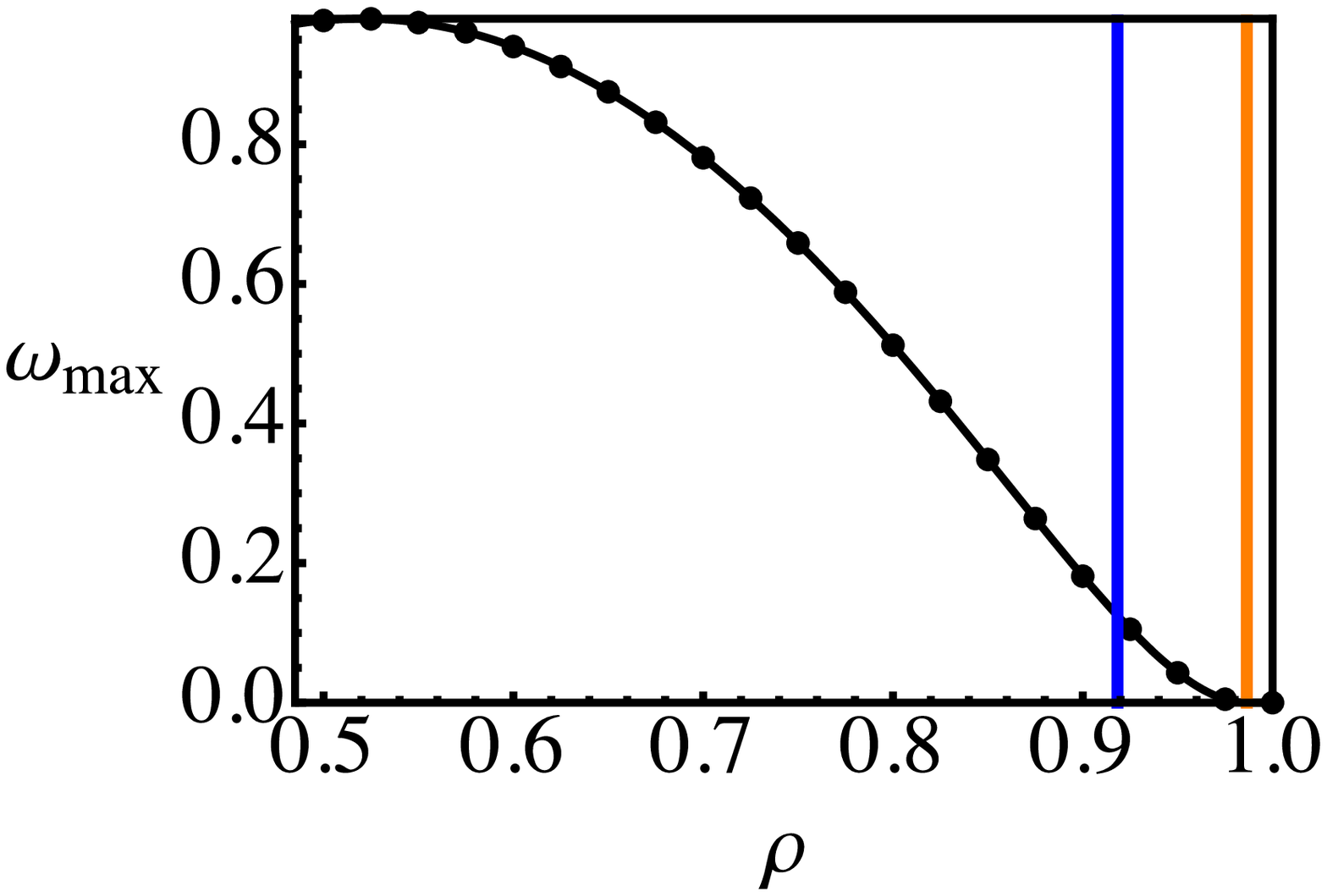}%
\label{wmax2a}}
\hspace{0.1cm}
\subfloat[]{\hspace{-0.25cm}\includegraphics[clip,width=0.5\columnwidth]{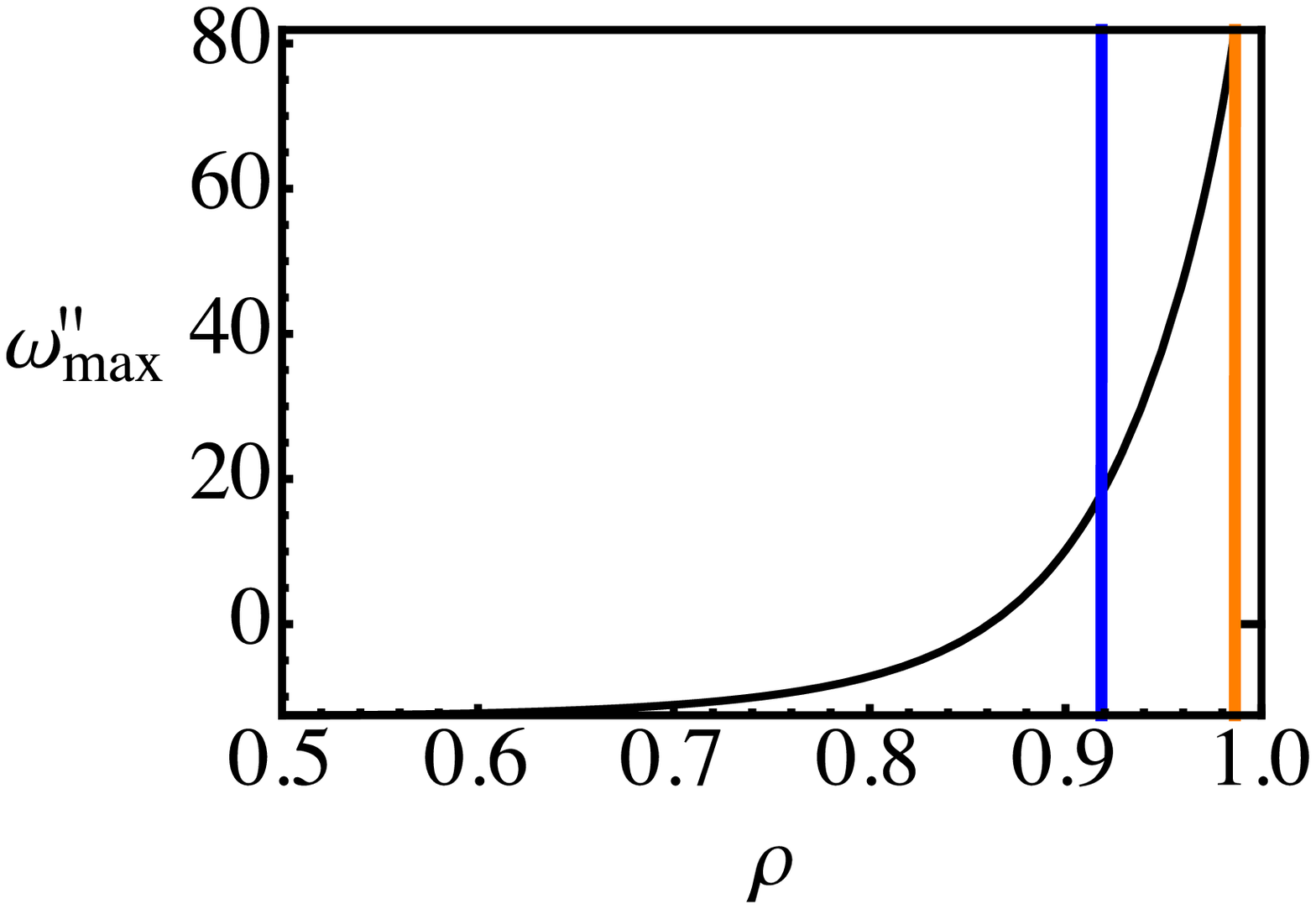}%
}
\caption{As Fig.\ \ref{wmax}, but for $w_s=0.5$. Note that the maximum in the second density derivative of $\omega_{\rm max}$ is now at the annealed spinodal density.}
 	\label{wmax2}
\end{figure}

In the monodisperse limit ($M=1$), expression (\ref{growthrates}) does not depend on $w_s$. This of course should be so as swaps between identical particles do not change the system configuration and hence cannot contribute to the dynamics. To find the maximum growth rate, one needs to maximize $\omega$ with respect to $\mathbf{k}$. As in the regime of interest the maximum occurs for small wavevectors, this can be done by expanding $\omega$ as a function of $\mathbf{k}$ around $|\mathbf{k}|=0$ up to the fourth order in $|\mathbf{k}|$, and then maximising the resulting expression analytically. One can also carry out numerically the maximization of the full $\omega(\mathbf{k})$, with essentially indistinguishable results (see Figures \ref{wmaxa} and \ref{wmax2a}), but we use the expansion procedure to obtain a closed-form expression for $\omega_{\rm max}$. (This is nonetheless too long to be displayed here, though.)
 
Because of the moment structure of (\ref{growthrates}), our linear theory can be applied to a fully polydisperse (i.e. $M \rightarrow \infty$) system, using an experimentally-reasonable distribution like the Schulz-Gamma form\cite{ratzsch1986cloud}
\begin{equation}
f^{(0)}(\sigma)= \frac{1}{g!}\left(\frac{g+1}{\bar{\sigma}}\right)^{g+1} \sigma^{g}\exp{\left[-\left(\frac{g+1}{\bar{\sigma}}\right)\sigma \right]}.
\label{schulz}
\end{equation}
In the following we set the mean interaction strength $\bar{\sigma}=1$ as we did in the binary case. The parameter $g$ controls the polydispersity of the distribution, which is given by
$1/\sqrt{g+1}$. This means that e.g.\ the choice $g=15$ produces a standard deviation of $\sigma$ that is  $25\%$ of the mean. With these choices, the moments appearing in the growth rates (\ref{growthrates})  can be expressed in terms of the density as
$\rho_1=\bar{\sigma}\rho=\rho$, $\rho_2=\bar{\sigma}^2\rho[1+1/(g+1)]=\rho(g+2)/(g+1)$. Note that because only moments up to second order appear in our mean-field spinodal rates, other distributions with the same mean and variance would give identical results.

Let us now see what our linear theory says about Warren's scenario as applied to the spinodal dynamics. 
Fig.\ \ref{wmaxa} shows the maximum growth rate as a function of the overall density, for reasonably dense systems. Here and in the following we fix the overall timescale 
by setting $w_0=1$.
The vertical lines indicate the upper limits of the annealed and quenched spinodal regions, respectively. 
As expected, the maximum growth rate becomes zero beyond the annealed spinodal, where the system is stable to all density fluctuations.
More remarkable is that the maximum growth rate increases only very slowly as density is decreased below this point, and only begins to rise substantially at the \textit{quenched} spinodal. This is exactly in line with what would be expected from Warren's two-stage hypothesis: inside the quenched spinodal region, the system has fast (stage 1) spinodal dynamics driven by the instability with respect to density fluctuations. (This corresponds to region R1 in Table~\ref{thetab}.) Outside the quenched spinodal, on the other hand, there is no spinodal decomposition in stage 1 (corresponding to regions R2/R3) and the spinodal dynamics is produced by the much slower growth of composition fluctuations in stage 2. To the extent that stage 2 dynamics, which involves local fractionation, is slow compared to stage 1, $\omega_{\rm max}$ should therefore be small between quenched and annealed spinodals, as compared to its values inside the quenched spinodal region. This is exactly what we find.

Graphically, the above reasoning means that $\omega_{\rm max}(\rho)$ should have a kink at the quenched spinodal, where it crosses over from small (stage 2) to large (stage 1) values.
The situation in Fig.\ \ref{wmaxa} is quite close to such an ideal two-stage scenario.
The kink can be seen more clearly by looking at the
%
%
\textit{second derivative} of $\omega_{\rm max}$ with respect to $\rho$, which would be large around a kink.
Fig.\ \ref{wmaxb} shows that this second derivative does indeed have a maximum, and this is positioned close to, if not exactly at, the quenched spinodal density.
Note that this happens even though our calculation of $\omega_{\rm max}$ did at no point involve any restriction to quenched dynamics, i.e.\ fixed composition. In other words, the full, unrestricted dynamics of the system nonetheless `feels' the presence of the quenched spinodal. 
This provides strong support for Warren's two-stage scenario.

\begin{figure}[h!]
	\centering
	\includegraphics[clip,width=\columnwidth]{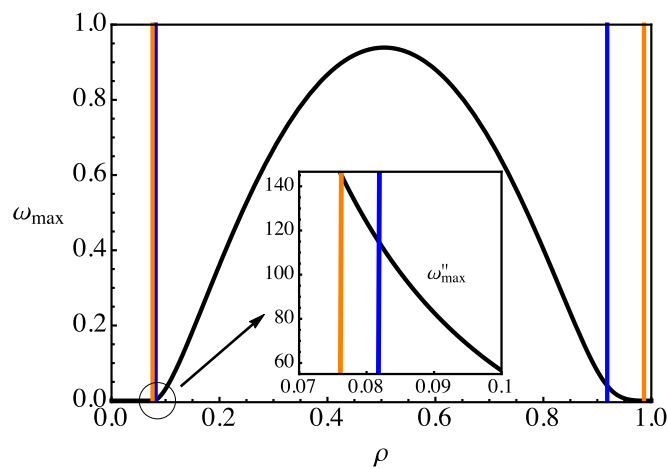}%
	\caption{Behaviour of $\omega_{\rm max}$ across the full density range, for the same parameters as in Fig.\ \ref{wmax}. The inset shows that the second derivative of $\omega_{\rm max}$ shows no special feature at the lower quenched spinodal density, where fractionation is too fast for Warren's two-stage scenario to apply.
}%
	\label{FullRangeRates}
\end{figure}

\begin{figure}[h!]
	\centering
	\includegraphics[clip,width=0.9\columnwidth]{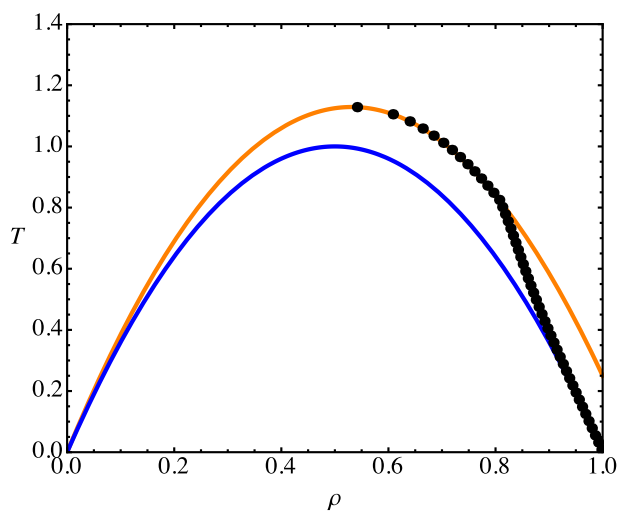}%
	\caption{Position of the maximum of the second derivative of $\omega_{\rm max}$, shown on the $x$-axis, versus temperature on the $y$-axis (points). 
Other parameters as in Fig.~\ref{wmax}. 
The curves give the quenched (lower curve) and annealed (upper curve) spinodals. Note that the second derivative maximum agrees closely with the quenched spinodal throughout the high density (above $\rho \approx 0.9$) region.}%
	\label{TvsRhoatmax}
\end{figure}

Since the physics of Warren's two-stage scenario requires the system to be \textit{dense}, we expect it to be break down at lower densities. To check this, we first investigate the behaviour of $\omega_{\rm max}\left(\rho\right)$ around the \textit{lower} end of the spinodal region. In Fig.\ \ref{FullRangeRates}, it is clear 
that the second derivative does not have any signatures around the (lower) quenched spinodal density, instead increasing smoothly towards the annealed spinodal.
Secondly, returning to the kink in $\omega_{\rm max}$ at the upper quenched spinodal density, we can consider its dependence on temperature. At  higher temperatures, the upper spinodal densities become lower, so that the two-stage scenario should be less pertinent.
To check this, we show in Fig.\ \ref{TvsRhoatmax} the density where the maximum in the second derivative of $\omega_{\rm max}$ for a range of temperatures. 
We can see that the density at the maximum moves away from the quenched spinodal curve and towards the annealed spinodal as $T$ increases, as expected. More usefully, we can read off from the figure that the two-stage scenario gives a good account of the position of the kink for densities above $\rho\approx 0.9$, so for the given polydispersity this is the threshold where the system becomes sufficiently dense to make fractionation slow. 

We now perform a second test that detects directly whether the behaviour we are seeing -- namely the near-kink in $\omega_{\rm max}(\rho)$ -- is in fact due to fractionation being slow. 
We do this by turning on direct particle swaps, using a nonzero swap attemp rate $w_s$. Fractionation is then possible even in dense systems, without relying on mediation by rare vacancies. 
The signatures of the two-stage scenario that we have found should therefore disappear as we increase $w_s$.
Indeed, Fig.\ \ref{wmax2} shows that for $w_s=0.5$, $\omega_{\rm max} $ increases smoothly as density is decreased from the annealed spinodal, rather than remaining small until the quenched spinodal. Likewise the  second derivative of $\omega_{\rm max}\left(\rho\right)$  is now featureless around the quenched spinodal and just increases gradually towards its value at the annealed spinodal. 
Essentially, this is evidence that the two-stage scenario has been destroyed. 

So far we have focussed on a system with fixed polydispersity of 25\%. As Warren's argument does not rely on specific features of the $\sigma$-distribution, one would however expect qualitatively similar results also for other polydispersities. To assess this quantitatively one needs to account for the fact that the separation in density between quenched and annealed spinodals grows with polydispersity. This can be done by considering the density difference between the maximum of $\omega_{\rm max}''$ and the quenched spinodal, normalized by the difference between the annealed and quenched spinodal densities. When this ratio is $\ll 1$, the kink of $\omega_{\rm max}$ is close to the quenched spinodal as the two-stage scenario predicts. Upon varying temperature, density and polydispsersity we find (data not shown) that the ratio is indeed dependent mostly on density and largely independent of polydispersity, becoming small at high densities as it should.

Summarizing the discussion in this section thus far, our mean-field theory for the spinodal dynamics of polydisperse colloids provides strong support for Warren's two-stage hypothesis, in the appropriate regime of high densities. It is worth noting that previous support for the two-stage scenario, both in Warren's original paper%
\cite{A809828J} and in the study by Pagonabarraga and Cates,\cite{ignacio} was obtained in the context of polymers, whereas here we have polydispersity in the interaction strength in a context that is more easily connected with the physics of colloids. This indicates that Warren's scenario may be a general feature of the non-equilibrium dynamics of dense fluid mixtures. 

We next briefly discuss the temperature dependence of the maximum growth rates, returning to the physical value $w_s=0$. The maximum growth rates are then directly proportional to $w_0$. So far we have taken constant $w_0=1$, though the results of Fig.\ \ref{TvsRhoatmax} would remain the same for any temperature-dependent $w_0$: this is because we always consider the second derivative $\omega_{\rm max}''$ with respect to density, at fixed temperature.
With constant $w_0$, we find that the typical maximum growth rates (say, in the middle of the spinodal region)  increase towards low temperatures as $\sim 1/T$.
This temperature dependence comes from the use of Glauber rates: linearizing the factors of $\exp(\Delta H/T)$, with energy changes from particle moves that are of temperature-independent magnitude, gives a factor of $1/T$ in all growth rates.
For quantitative modelling of spinodal growth rates one would therefore want to choose a $w_0$ that goes to zero as $T\to 0$, e.g.\ $w_0\sim T$, which would be consistent with the picture of an underlying diffusive dynamics causing attempted particle moves.
This would then give constant maximum growth rates for low $T$.

We conclude this section with a discussion of the amplitudes $\delta p^{\alpha}$ of the mode that grows fastest in the spinodal dynamics, with rate $\omega_{\rm max}$. These amplitudes define the relative strength of density fluctuations for each species. As a vector in density space, they identify the early-time (non-equilibrium) fractionation direction, or simply \textit{spinodal direction}.
As discussed above, combining eqns (\ref{LinearDMFT}) and (\ref{planewave}) leads to an eigenvalue equation with eigenvalue $\omega$, and the $\delta p^{\alpha}$ are the $M$ components of the eigenvector corresponding to $\omega_{\rm max}$.
This eigenvector can be calculated by solving the eigenvalue equation numerically, or one can derive a closed-form expression in terms of $\omega$, $\rho$, and the attempt rates. We will denote the angle between the spinodal direction and the dilution line by $\theta_{\rm max}$. This quantity is of interest because in an ideal two-stage scenario, it should be zero inside the quenched spinodal region, where the initial dynamics corresponds to stage 1, i.e.\ pure density fluctuations; it should then rise as one moves from the quenched spinodal density to the annealed one, where the spinodal dynamics corresponds to stage 2. Our numerical data for $\theta_{\rm max}(\rho)$ (not shown) follow this scenario fairly closely in dense systems, providing further support for Warren's hypothesis. As our system is not ideal in the sense that fractionation is not infinitely slow, we find inside the quenched spinodal region a $\theta_{\rm max}$ that is constant but not quite zero, indicating that even in stage 1 the dominant spinodal mode contains a fractionating component in addition to pure density fluctuations. Likewise the transition to larger values of $\theta_{\rm max}$ is not a sharp kink but a crossover, though importantly this remains located around the quenched spinodal density. As in the case of $\omega_{\rm max}$, we have tested that when direct particle swaps ($w_s>0$) are introduced, this crossover disappears and is replaced by a featureless increase of $\theta_{\rm max}$ across the entire annealed spinodal region. This confirms that the crossover in the physical system ($w_s=0$) arises from fractionation  being a slow process.

\section{Beyond the spinodal regime}
\label{numerics}
In order to see what happens after the end of the spinodal regime, and hence investigate the full phase-separation dynamics, we integrate our mean-field kinetic eqns (\ref{kineticequations}) numerically using a forward Euler method, with periodic boundary conditions. The method returns the evolution of the local densities for all species. For ease of visualization and reduction of computation time, spatial dimension $D=2$ was chosen. Firstly, we examine what happens for binary mixtures ($M=2$), and then extend the analysis to $M>2$. We set the jump attempt rates as before, i.e.\ $w^{\alpha0}=w_0$ and $w^{\alpha\beta}=w_s$, for any $\alpha\neq0$ and $\beta\neq0$. More specifically we concentrate on the physical setting $w_s=0$, unless otherwise stated. To find the evolution of a system whose overall composition is given by a list of densities $\left\lbrace p^{\alpha} \mid \alpha=1,\dots,M \right\rbrace$, we firstly created an initial homogeneous state defined by $p^\alpha_i=p^\alpha$ for all sites $i$.
To trigger the phase separation, we added small fluctuations to the initial state of each species by generating $L^D$ random numbers, normally distributed, with mean zero and standard deviation $1\%$. We then subtracted the average of these random numbers, to ensure that the overall density of every species remains unchanged. The time step used was limited by the numerical stability of the algorithm, and typically equal to $0.1$; we checked that this value is small enough to give us effectively the solution of the continuous-time equations by running the numerics for exactly the same initial configuration with a time step smaller by a factor of 5 and verifying that the results were virtually the same. When we used $50\times50$ lattice sites and $M=2$, for instance, our program (written in C) performed $8,000$ time steps in approximately $100$ seconds (running at 2.6 GHz processor speed), which in many instances was enough time to grow relatively large domains.

\subsection{Binary fluids}
\label{binumerics}
\begin{figure}[h]
	\centering
	\setbox1=\hbox{\includegraphics[width=\columnwidth]{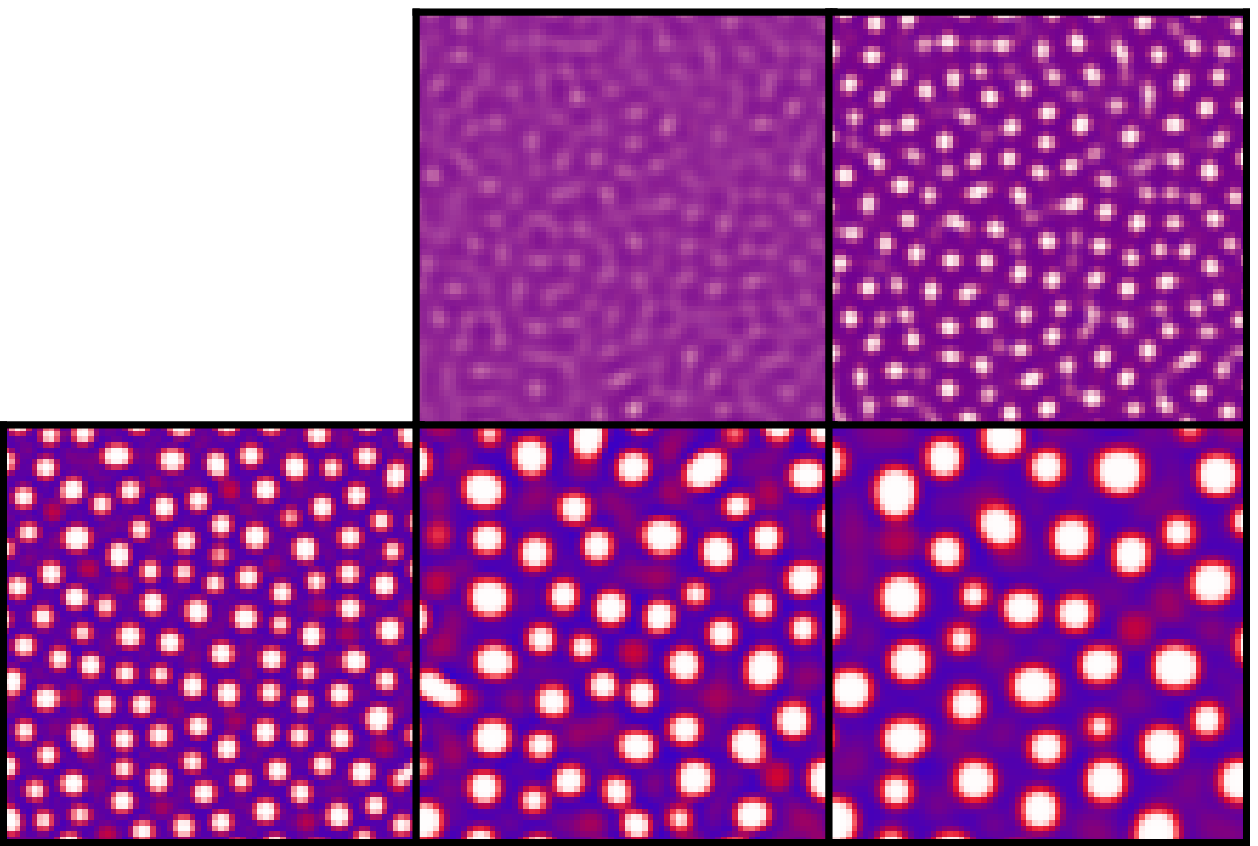}}
	\includegraphics[width=\columnwidth]{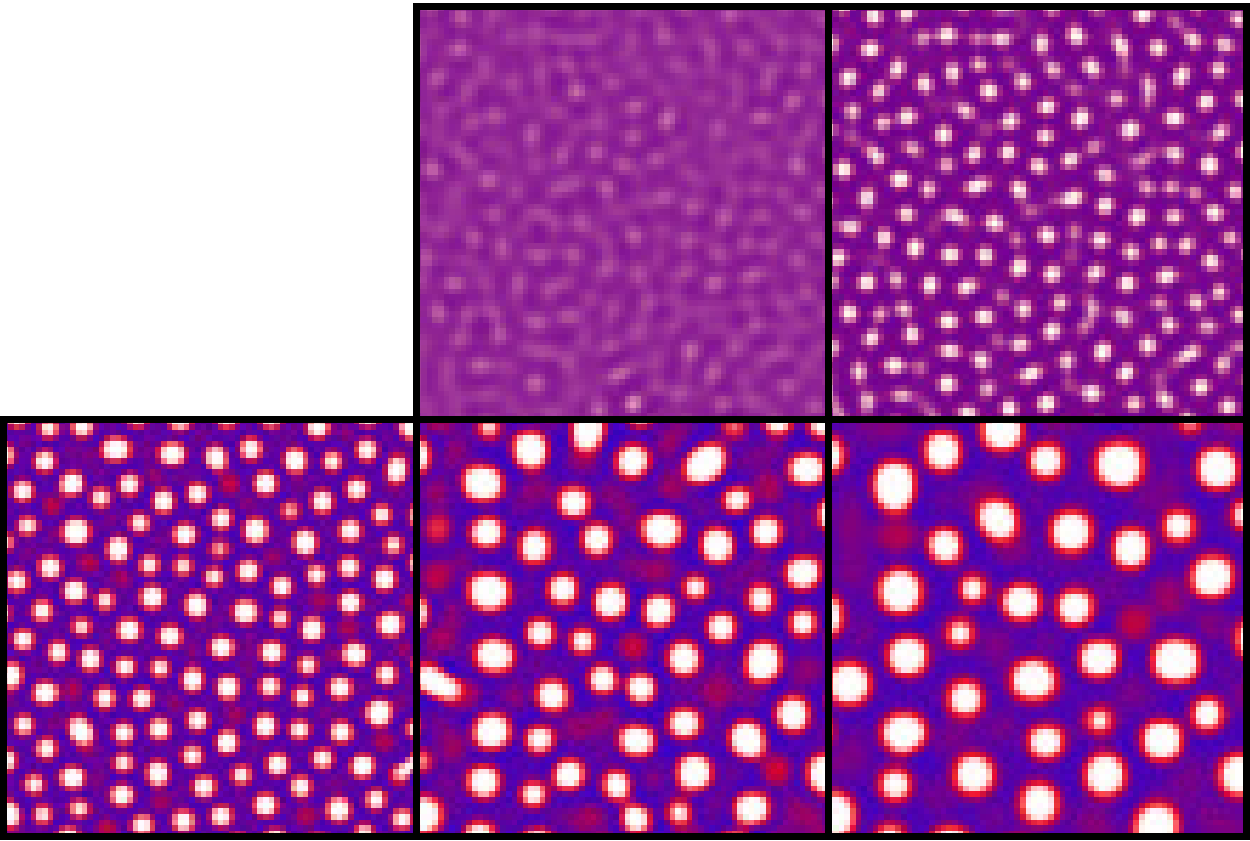}\llap{\makebox[\wd1][l]{\raisebox{3.2cm}{\hspace*{-0.1cm}\includegraphics[width=0.95\columnwidth/3]{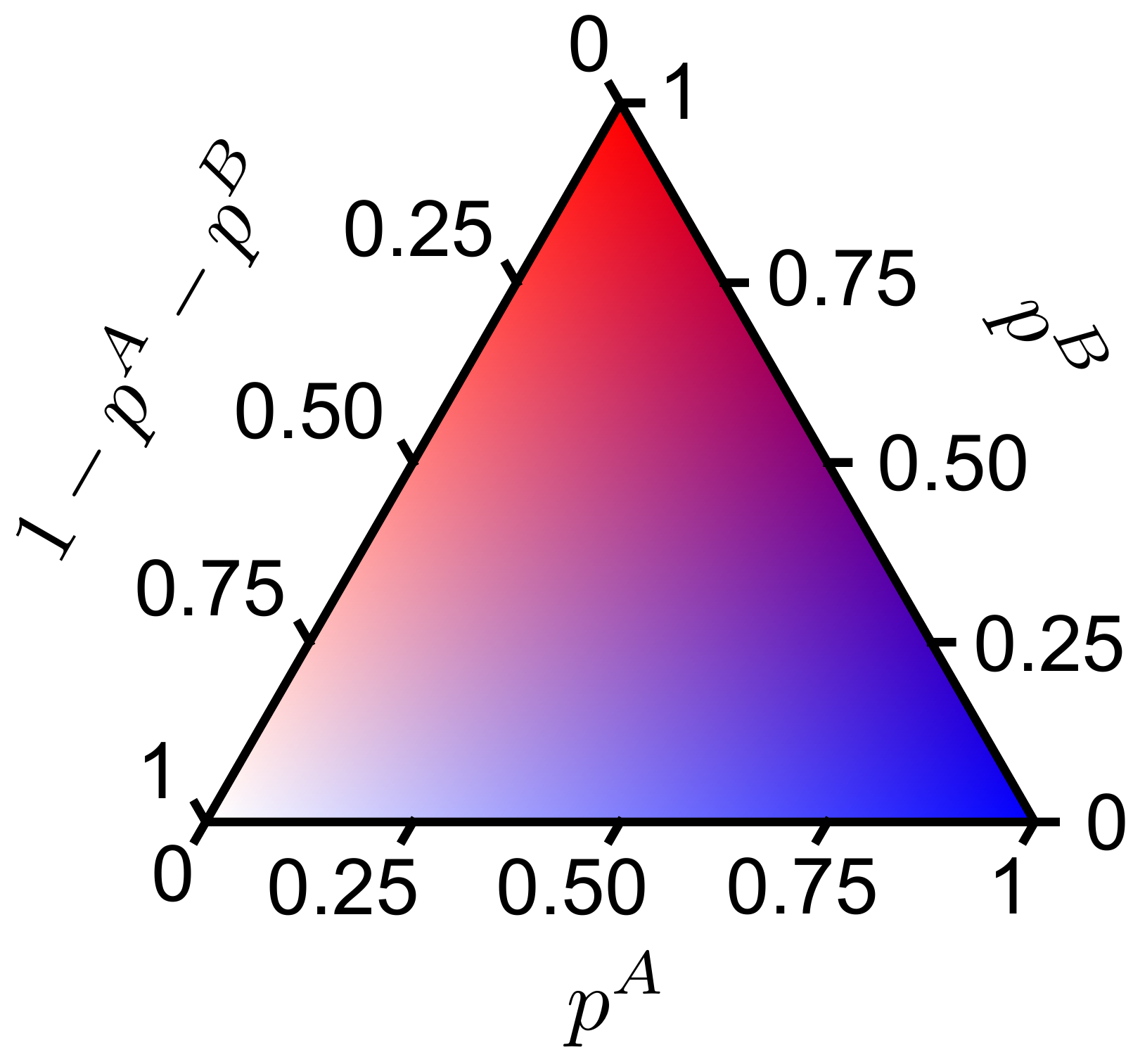}}}}
	\caption{Time snapshots showing the local compositions throughout the system, as from numerics. The colour scheme is based on the RGB colour model, leading to the colour key shown in the top-left corner; see text for details. The lighter portions of the system are gas bubbles, which are surrounded by a $B$-rich interface separating them from an $A$-rich continuous liquid phase. Parameters: $p^A=p^B=0.41$, $T=0.3$, $d=0.25$, $w_s=0$, and $L=150$. (Only a $75\times75$ region of the system is shown here.) From top centre to bottom right, the snapshots are taken at $t=8$, $16$, $316$, $4850$ and $16200$.}
	\label{snapshots_d05}
\end{figure}

Fig.\ \ref{snapshots_d05} shows snapshots of the phase-separation dynamics for $M=2$, with $\sigma_A=1+d$ and $\sigma_B=1-d$. For these binary case numerics we chose $d=0.25$. Thus the first and second moment densities (normalized by $\rho$), i.e.\ $\rho_1/\rho$ and $\rho_2/\rho$, are the same as in the continuous distribution analysis presented in Section \ref{linear}, where the Schulz parameter was $g=15$; also, the same temperature $T=0.3$ was used. Therefore the spinodal growth rates 
are also the same as in the fully polydisperse case and we expect, 
at a given density $\rho$, to see a very similar initial dynamics.
We chose $\rho=0.82$, which places the system within the quenched spinodal region. To determine the colour of a site $i$, we used a colour scheme in which the colours red, green, and blue are blended together. The intensity of each of these colours at a given site varies from $0$ to $1$. In our scheme, red, green, and blue intensities are given exactly by $1-p_i^A,p_i^0,1-p_i^B$, respectively. (Remember the notation for the local concentration of vacancies, i.e.\ $p_i^0=1-p_i^A-p_i^B$.) This leads to the colour key shown in the top-left part of Fig.\ \ref{snapshots_d05}. It is plotted in triangular colour space in $(p^A, p^B, 1 - p^A - p^B)$, dropping the site index $i$. For example, if the concentration of particles of species $A$ at one site is high (low), and the concentration of particles of species $B$ at the same site is low (high), then the site colour will tend towards blue (red); if the concentrations of all species are all low, then the site colour will tend to white.

The snapshots in Fig.\ \ref{snapshots_d05} show the growth of lighter regions of the system that represent gas bubbles. These bubbles are surrounded by a $B$-rich (red) interface separating them from an $A$-rich (blue) continuous liquid phase. Walking from the centre of a gas bubble along an arbitrary direction, one therefore initially sees low concentrations of both particle types, then a high concentration of the particle species with the smallest $\sigma$, i.e.\ $B$-particles, and eventually one reaches the bulk liquid that contains predominantly $A$-particles, for which $\sigma$ is the largest. (Video showing the full-time evolution of the phase separation process is provided as an animation in the ESI.\dag)

Of course as we decrease $\rho$ our numerical results show larger vapour fractions. Changing $\rho$ can also lead to more complicated morphologies such as in Fig.\ \ref{illustrative}, which was generated using $T=0.4$, $p^A=p^B=0.33$, $d=0.3$, $w_s=0$, $L=128$, with the last snapshot taken at $t=6000$. Because the value of $\rho$ here is closer to the critical density one observes bicontinuous domains of gas and liquid. The intuition is the same as in a 
one-species Ising lattice-gas system, where because of the particle-hole (vacancy) symmetry, in a system at the critical density neither gas or liquid can `win' to form
bubbles or droplets.
Instead, finger-like bicontinuous structures are formed. Moving away from the critical point, one expects these to survive for a certain time until the system `notices' which phase is going to be the majority phase, and forms bubbles or droplets of the minority phase.

\begin{figure}
	\centering	
	\subfloat[\hspace{0.3\columnwidth} \normalsize$t=8$]{%
		\includegraphics[clip,width=0.8\columnwidth]{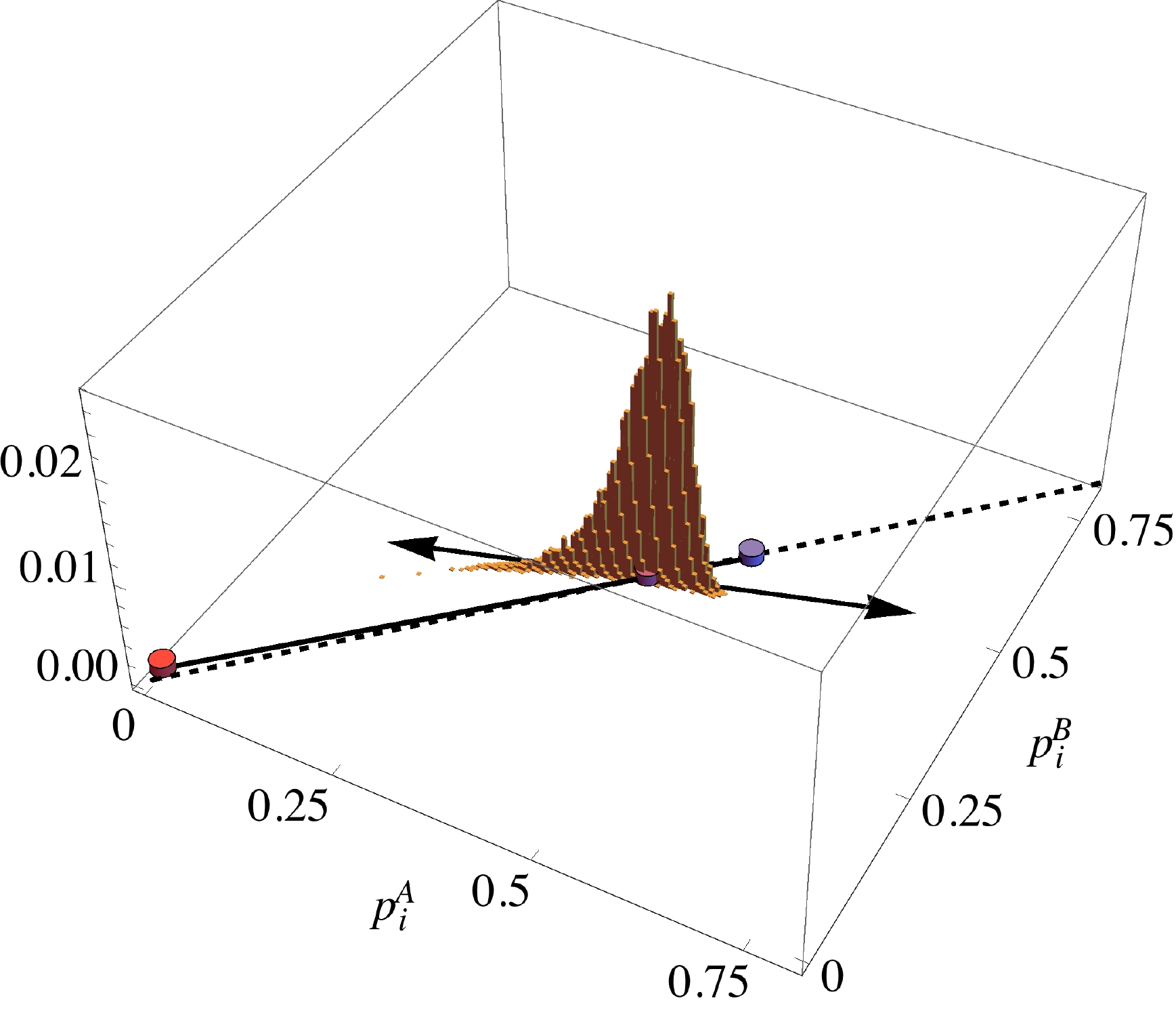}%
		\label{figu1}
	}
	
	\subfloat[\hspace{0.3\columnwidth} \normalsize$t=16$]{%
		\includegraphics[clip,width=0.8\columnwidth]{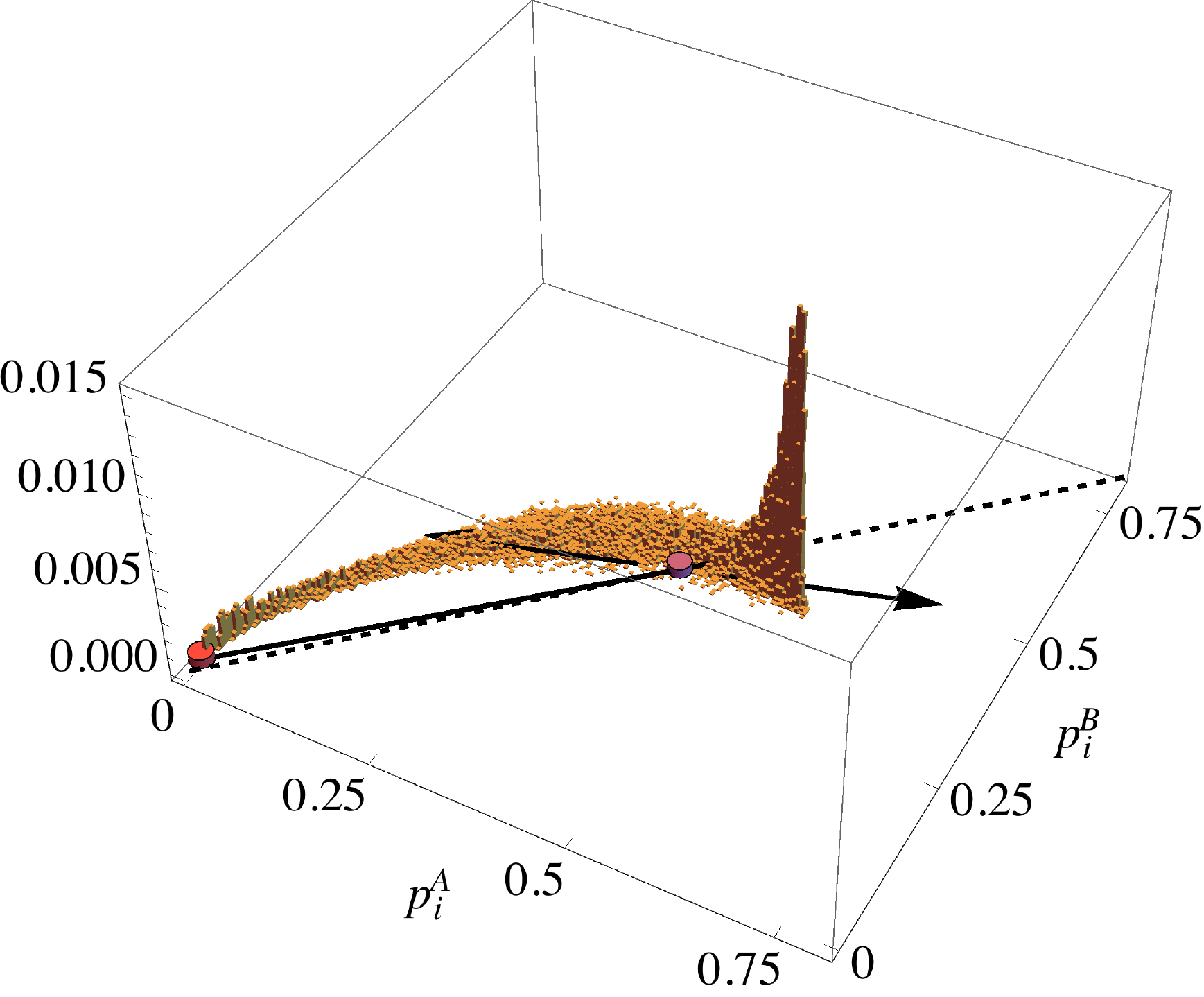}%
		\label{figu2}
	}
	
	\subfloat[\hspace{0.3\columnwidth} \normalsize$t=316$]{%
		\includegraphics[clip,width=0.8\columnwidth]{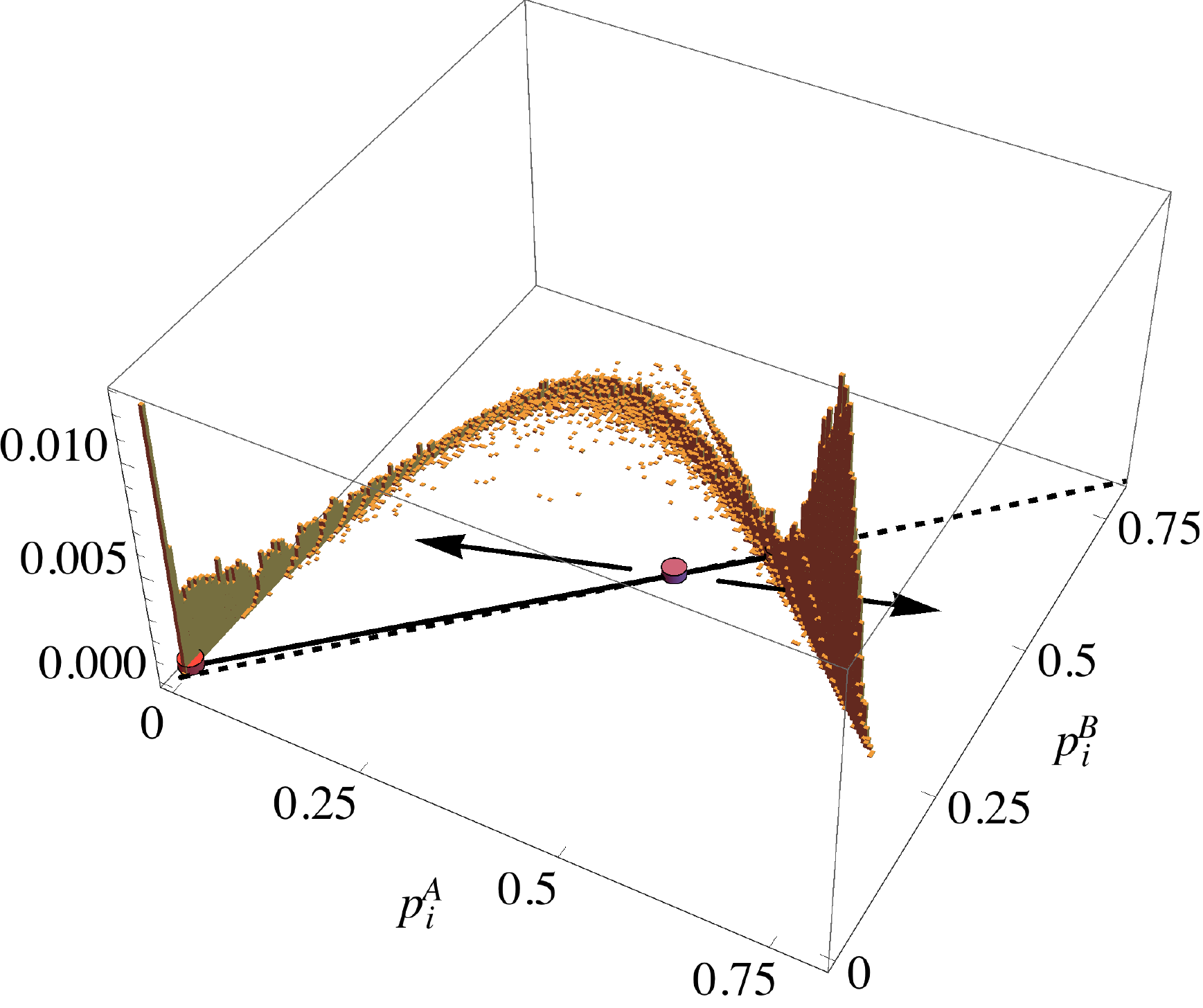}%
		\label{figu3}
	}
	\caption{Time evolution of two-dimensional density histograms, in the $(p^A_i,p^B_i)$-plane, for the data used in Fig.\ \ref{snapshots_d05}; see text for details. The bin width along each dimension is $0.005$. The parent phase lies on the dilution line (dashed). The red and blue dots off the dilution line mark the low- and high-density daughter phases obtained from our equilibrium numerics. The connecting tieline contains the parent and defines 
the equilibrium fractionation direction. The spinodal direction, which was obtained from our early-time analysis, is shown by the double-headed arrow.}
	\label{histo}
\end{figure}

In Fig.\ \ref{histo} we introduce a different representation of the time evolution that will prove to be quite revealing. We show two-dimensional \textit{density histograms}, in species density space, i.e.\ in the $p^A_i$--$p^B_i$ plane that contains all the possible density combinations an arbitrary site $i$ can have; the physically accessible region in this plane is a triangle bounded by 
$p^A_i\geq 0$, $p^B_i\geq 0$ and $p^A_i+p^B_i \leq 1$). What the histogram counts is the number of lattice sites $i$ that have species densities $(p^A_i,p^B_i)$ inside each two-dimensional bin. The results were normalized by the total number of lattice sites $L^D$. Such a density histogram can then be viewed as a dynamic analogue of an equilibrium phase diagram as sketched in Fig.\ \ref{fracbi} above.
In a density histogram, the parent phase lies on the dilution line (shown dashed in Fig.\ \ref{histo}) as before. The low- and high-density daughter phases calculated from the equilibrium phase diagram lie off this dilution line, with the parent on the connecting tieline; the latter defines the equilibrium fractionation direction. At the  temperature we are considering, this equilibrium fractionation direction deviates only slightly from the dilution line. From the histograms, we can clearly see different dynamical regimes: initially, the histogram spreads linearly from the parent along the spinodal direction as expected for spinodal dynamics. As nonlinear effects kick in, a curved path of compositions connecting a gas and liquid phase is then formed. This clearly delineated `arc' contains the compositions of the different parts of the system: as one moves in space from a gas bubble into the bulk liquid, one passes through a series of compositions within the interface between these two phases. 

Beyond this generic structure, there are several interesting observations we can make from Fig.\ \ref{histo}. The density histograms reveal that gas-liquid interfaces are strongly fractionated, with the arc being well away from the dilution line, in the $B$-rich part of the density plane. Physically, the reason is that $B$-particles have smaller $\sigma_B$ and hence interact more weakly; they therefore pay a smaller energy penalty for sitting at an interface, where they have fewer neighbouring particles. Interfaces also have a well-defined sequence of density combinations as can be seen from the fact that the gas-liquid arc is quite narrow.

\begin{figure}[h]
	\centering
	\includegraphics[width=0.9\columnwidth]{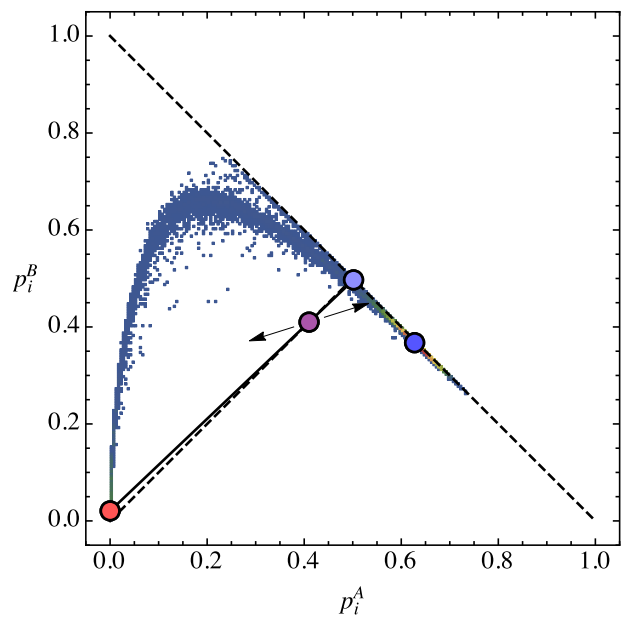}
	\caption{Same histogram as in Fig.\ \ref{figu3}, but now as seen from the top; different heights are represented by different colours. Additionally, we show the high-density peak of the histogram, which gives the majority composition of the bulk liquid.
}
	\label{denshistOVERFRAC}
\end{figure}

For further analysis it is useful to switch to the two-dimensional density histogram representation in Fig.\ \ref{denshistOVERFRAC}. 
This shows the same data as in Fig.\ \ref{figu3} but now seen from the top, with different heights corresponding to different colours. The peak in the high-density region of the histogram is also marked; this gives the majority composition of the bulk liquid
at that time instant. It is interesting to observe that this peak, having started out at the parent composition and `walked' along the spinodal direction initially, does not subsequently move straight away towards the liquid equilibrium composition. Instead, the liquid phase composition stays away from its equilibrium optimum for a long time. In fact, for the dynamics shown in Fig.\ \ref{denshistOVERFRAC} the liquid peak moves \textit{away} from its equilibrium point for a long transient period beyond the initial spinodal decomposition dynamics. This 
arises because the gas-liquid interfaces are strongly enriched in $B$-particles, leaving an unusually $A$-rich bulk liquid.
Of course at very long times the equilibrium prediction and the dynamics must eventually agree, and we have verified that the density histogram peaks then indeed centre on the calculated equilibrium compositions while
the arc with the interface compositions contains only a small ($\sim W/L$, where $W$ is the interfacial width) fraction of probability. 
(See also the evolution of two-dimensional density histograms in the animation in the ESI.\dag)

\begin{figure}[h!]
	\centering	
	\subfloat[]{%
		\includegraphics[width=0.9\columnwidth]{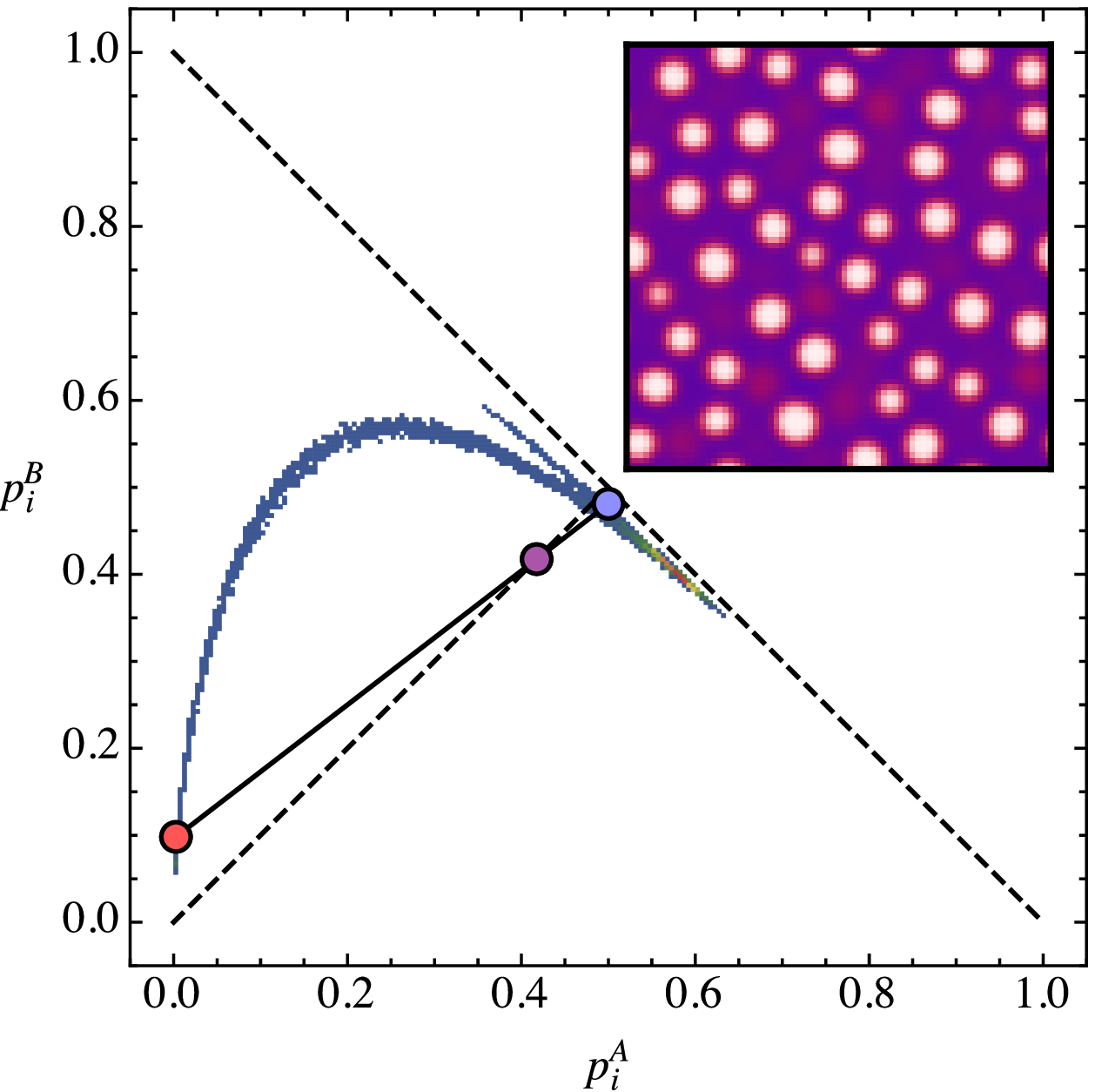}%
		\label{arm}
	}
	
	\subfloat[]{%
		\includegraphics[width=0.9\columnwidth]{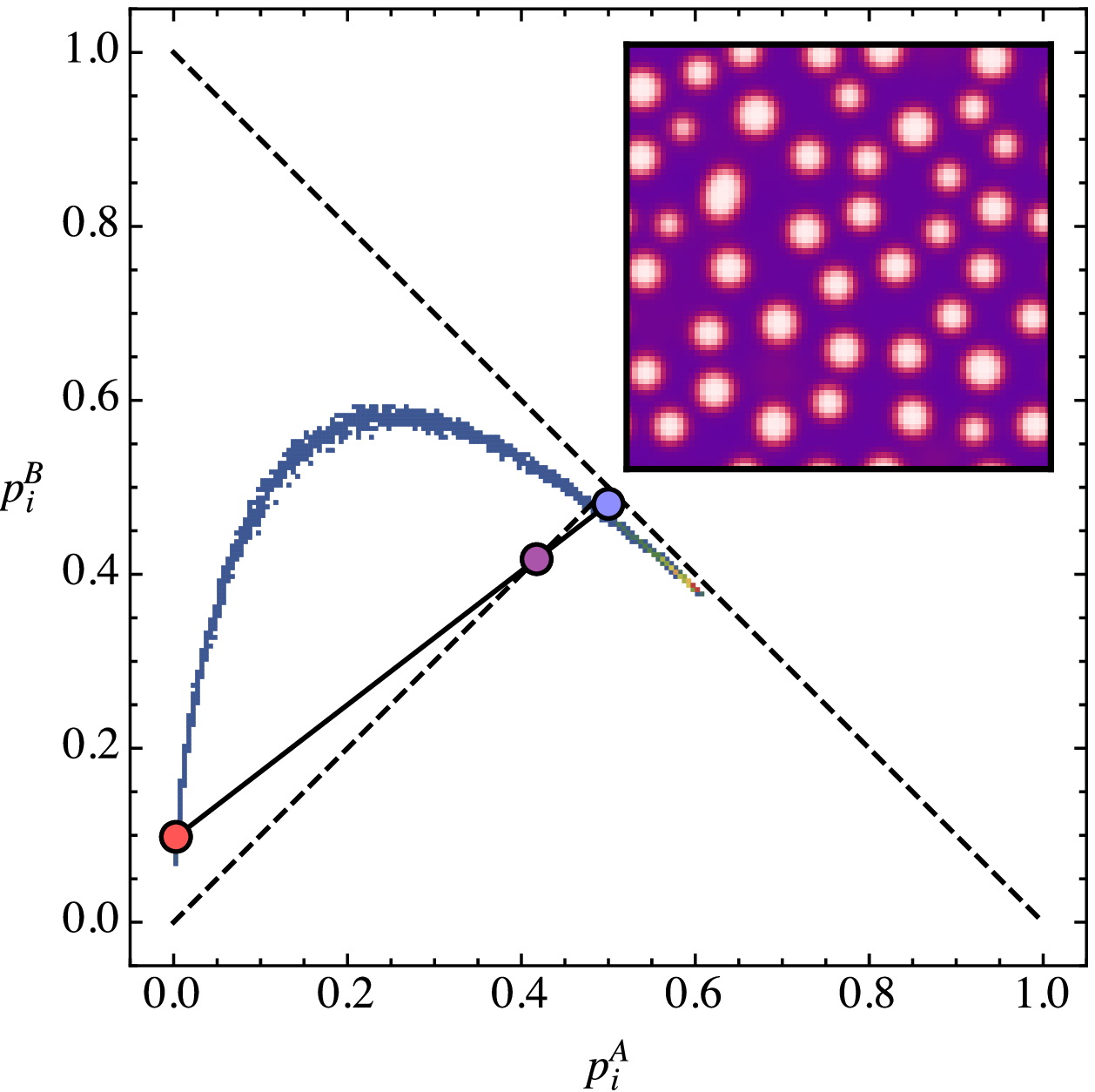}%
		\label{arm2}
	}
	\caption{(a) Density histogram showing a clear `arm', a composition heterogeneity in the bulk liquid. This feature arises from the evaporation of gas bubbles: the remnants of their interfaces rapidly equilibrate to the liquid density but only slowly relax their composition, producing the $B$-rich patches (red) visible in the inset. Parameters: $p^A=p^B=0.4175$, $T=0.5$, $d=0.25$, $w_s=0$, and $L=75$, at $t=1110$. The inset shows the corresponding snapshot in real space. (b) Same as (a), but now $w_s=0.5$.}
\end{figure}

A final, intriguing feature of the dynamics we observe is the inhomogeneity of the bulk liquid: in the last two snapshots in Fig.\ \ref{snapshots_d05} one can clearly see well-defined liquid regions that are unusually enriched in $B$-particles. In the density histograms of Figs.~\ref{figu3} and \ref{denshistOVERFRAC} and also Fig.\ \ref{arm}, which shows results at a higher temperature, these regions manifest themselves as an `arm' at high density that is quite distinct from the arc arising from gas-liquid interfaces.

Looking at Fig.\ \ref{snapshots_d05}
carefully, one notices that the origin of the $B$-rich liquid regions lies in the evaporation of gas bubbles. As these shrink, so do their interfaces, eventually forming dense patches. Because the interfaces are strongly fractionated, these dense  patches are strongly enriched in $B$-particles. While the density of the patches can rapidly equilibrate to the bulk liquid\textemdash as shown by the fact that the arm almost coincides with a line of constant total number density $p^A+p^B$, as seen from Figs.\ \ref{figu3}, \ref{denshistOVERFRAC}, and \ref{arm}\textemdash it requires inter-diffusion of particles to equilibrate their composition. Hence the composition heterogeneities formed by these patchese are unusually long-lived. We thus have here, in the long-time dynamics, another striking manifestation of Warren's hypothesis, in its general form which says that equilibration of composition is slow in dense systems.
In order to check 
this interpretation
we turned on $w_s$; as expected, the arm then disappears, and in the real-space images the liquid is clearly homogeneous (Fig.\ \ref{arm2}).

\subsection{Multicomponent fluids}
Finally, we show now that a similar two-dimensional density histogram analysis can be performed even when one considers \textit{arbitrary} $M$. A full density histogram in $(p_i^1,\ldots,p_i^M)$-density space would be $M$-dimensional, so one needs to project down to a manageable low-dimensional representation.
The obvious choice for the new histogram axes are (the local versions of) low-order moment densities, 
which can be used for arbitrary $M$. 
We choose 
the local analogues of $\bar{\sigma}\rho_0-\rho_1$ and $\rho_0$. The reason for this choice is that in the bidisperse case ($M=2$),
\begin{equation}
\bar{\sigma}\rho_0-\rho_1=
(\bar{\sigma}-\sigma_A)p^A+
(\bar{\sigma}-\sigma_B)p^B
=d(p^B-p^A)
\end{equation}
while $\rho_0=p^A+p^B$, so that the new histogram axes are just a rotated version (by 45$^\circ$) of the ones we have used so far.

\begin{figure}[h]
\centering
\scalebox{1.05}{
\begin{tikzpicture}
\hspace{-0.3cm}
	\node (img1) {\includegraphics[width=1.0\columnwidth]{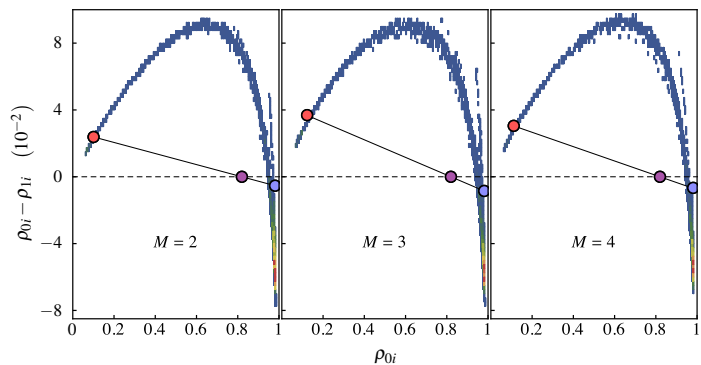}};
	\coordinate (x) at (-4.15,1.675);
	\node[font=\color{black},scale=0.7] at (x) {$8$};
	\coordinate (x) at (-4.15,1.675-0.85);
	\node[font=\color{black},scale=0.7] at (x) {$4$};
	\coordinate (x) at (-4.15,1.675-0.85-0.85);
	\node[font=\color{black},scale=0.7] at (x) {$0$};
	\coordinate (x) at (-4.15-0.09,1.675-0.85-0.85-0.85);
	\node[font=\color{black},scale=0.7] at (x) {$-4$};
	\coordinate (x) at (-4.15-0.09,1.675-0.85-0.85-0.85-0.85);
	\node[font=\color{black},scale=0.7] at (x) {$-8$};
	\coordinate (x) at (-5.7,1.675-0.85-0.85+0.05);
	\node[rotate=90, anchor=center,yshift=-1.1cm,font=\color{black},scale=0.75]  at (x) {$\rho_{0i}-\rho_{1i} \hspace{0.25cm}\left(10^{-2}\right)$};
	\node[below=of img1, node distance=0cm, yshift=1.1cm,font=\color{black},scale=0.75] {$\rho_{0i}$};
	\coordinate (x) at (-4.15-0.09+0.27,-1.965);
	\node[font=\color{black},scale=0.7] at (x) {$0$};
\end{tikzpicture}}
	\caption{Density histogram for different number of species $M=2,3,4$. The set of moment densities $\rho_0$, $\rho_1$, and $\rho_2$ is the same in all three cases. (This was achieved by appropriate choice of the parameter $d$, which was set to $0.25$, $0.354$, and $0.354$, respectively; see text for details.) The other parameters are $\rho=0.82$, $T=0.5$, $w_s=0$, and $L=50$, and $t=760$. The bin widths along the axes $\rho_{0i}$ and $\rho_{0i}-\rho_{1i}$ are $0.005$ and $0.0025$, respectively.}
	\label{Ms}	
\end{figure}

One can now use these quantities to plot two-dimensional density histograms for an arbitrary polydisperse system. It turns out that all the features that had been previously observed in the $(p^A,p^B)$-plane density histogram (i.e.\  the clearly delineated curved arc of interfaces compositions, the `arm', etc.)~remain qualitatively identical in this new coordinate system. Fig.\ \ref{Ms} shows our results for systems with two, three, and four species. For $M=3$, we used $\sigma \in \lbrace1-d,1,1+d\rbrace$, with relative densities (composition) given by $\lbrace\frac{1}{4},\frac{1}{2},\frac{1}{4}\rbrace$, whereas for $M=4$ we chose $\sigma \in \lbrace1-d,1-d/2,1+d/2,1+d\rbrace$ with composition $\lbrace\frac{1}{6},\frac{1}{3},\frac{1}{3},\frac{1}{6}\rbrace$. We then identified values of $d$ for each $M$ that give the same set of moment densities ($\rho_0$, $\rho_1$, and $\rho_2$) as we had in Section \ref{binumerics}. (Thus the spinodal growth rates are also all the same and hence results at the same $t$ are comparable\textemdash but similar results were found using different parameter sets chosen by the same method.) Even though these histograms are now projections of $M$-dimensional histograms to two dimensions, we still get thin arcs between daughter phases, showing that there is still a well-defined sequence of compositions in the interfaces.
We have performed checks over a larger parameter range, where we find that long-lived composition heterogeneities also appear in situations where one expects a pronounced slowing down of fractionation, exactly as we saw for $M=2$; correspondingly, they disappear (data not shown) when direct particle swaps are turned on.

The fact that the density histograms for different $M$ in Fig.\ \ref{Ms} are so similar is quite remarkable. This similarity is surprising because the mean-field dynamical equations (\ref{kineticequations}) do not in general reduce to closed dynamical equations for local moments like $\rho_0$ and $\rho_1$, due to nonlinear dependences on particle size $\sigma$ in the Glauber rates. Nonetheless our numerical results suggest that degrees of freedom not captured by these moments only influence the dynamics weakly, so that they could provide a useful way of thinking about the dynamics even for $M\to \infty$.

\section{Conclusions}
\label{conclusion}
In this work we investigated the dynamics of how colloidal polydisperse systems phase-separate, by introducing new kinetic equations based on the Polydisperse Lattice-Gas model.\cite{PhysRevE.77.011501} As a baseline for our mean-field approach to the dynamics we calculated the mean-field equilibrium phase diagram of the model, including cloud and shadow curves as well as 
spinodals.
To test Warren's two-stage scenario,\cite{A809828J} we obtained both the annealed and quenched versions of the phase diagram, the latter referring to a system that can only change its density but has fixed composition.
We analysed the linearized dynamical equations to understand the dynamics of spinodal decomposition, and found clear evidence in support of Warren's proposal. For the late-stage dynamics, we introduced a two-dimensional density histogram method that allows fractionation effects in the phase-separation dynamics to be clearly visualized. This revealed strongly fractionated interfaces between gas and liquid. It also helped us to detect the existence of long-lived composition heterogeneities in the bulk liquid, which are a further manifestation of the fact that fractionation dynamics is slow as suggested by Warren. This prediction may be amenable to experimental verification in dense colloidal systems. 
The whole analysis was performed for an arbitrary number of particle species, although much of it was presented in binary mixture context for the sake of simplicity and ease of visualization.

Our main assumptions were that the dynamics can be described by a kinetic lattice model, and that a mean-field approximation is at least qualitatively accurate.
One could ask whether non-mean-field effects in the equilibrium phase diagram of the PLG might cause significant changes in the phase-separation dynamics. This could be tested by deploying higher order approximations beyond our dynamical mean-field theory (DMFT), such as the Path Probability Method (PPM).\cite{gouyet2003description} One might expect that the PPM would not necessarily lead to new qualitative outcomes for the analysis presented here, since no differences were observed in a previous comparison between PPM and DMFT, though in the somewhat different context of relaxation dynamics in porous materials.\cite{edison2012modeling} Direct Kinetic Monte Carlo (KMC) simulations could be used to directly probe the dynamics of the PLG model and so assess the effects of our mean-field approximation. Such simulations were performed in ref.\ \citenum{tafa2001kinetics} for similar systems consisting of two species plus vacancies, where $p^A=p^B$ and the Hamiltonian is given by
\begin{equation}
H = - \sum\limits_{\langle i,j\rangle}^{}\sum\limits_{\alpha, \beta}^{}\epsilon_{\alpha\beta}n_{i}^{\alpha}n_{j}^{\beta}
\end{equation}
with $\alpha=A,B$. This setting includes the PLG Hamiltonian (for $M=2$) given by eqn (\ref{plg}) but allows more generic interactions $\epsilon_{\alpha\beta}$. Also using a mean-field phase diagram as a reference, the authors of ref.\ \citenum{tafa2001kinetics}
investigated segregation kinetics with particle-particle swaps forbidden (corresponding to $w_s=0$ in our notation) or highly energetically suppressed. They focussed on the existence of different domain growth morphologies in various parameter regimes. 
However, their choices for the interaction strengths $\epsilon_{\alpha\beta}$ were rather different from ours. Our separable assignment $\epsilon_{\alpha\beta}=\sigma_\alpha \sigma_\beta$ makes liquid-gas phase separation the dominant physical process, for which we then study the effects of polydispersity.
On the other hand, the authors of ref.\ \citenum{tafa2001kinetics} look at much less attractive (and even repulsive) $AB$ interactions, always with $\epsilon_{AA}=\epsilon_{BB}$. This leads to distinct physical effects, including the condensation of vacancies at interfaces between $A$- and $B$-rich phases.

KMC simulations could also be used to fit our model parameters to experimental systems. Ref.\ \citenum{edison2009modeling} explains that one way of obtaining a physically reasonable value for the particle-vacancy jump attempt rate $w_0$ is by comparing between estimated ($\hat{D}$) and experimental ($D_s$) self-diffusion coefficients. If $\hat{D}$ is obtained from KMC simulations in dimensionless units, 
the desired correspondence would be $D_s = \hat{D}a^2w_0$, where $a$ is the lattice spacing. Therefore, fixing the value of $a$, an estimate of $w_0$ can be obtained. (In principle, one could try to develop a similar scheme to obtain a value for $w_s$.) 

Coming to the limitations of the PLG model itself, one obvious shortcoming is the fact that a lattice model cannot capture gradual increases in density that are possible in an off-lattice setting, where even in an already fairly dense system collective motion can reduce the typical distance between particles. In a lattice model, on the other hand, density increases have to come from localized filling in of vacancies, or equivalently vacancies diffusing away. As we explain below the presence of vacancies also enables fractionation. Because both relaxation of density and of composition are then tied to the presence of vacancies, it is clear that the latter cannot become arbitrarily slow compared to the former. In a continuum model, however, one would expect that large pockets of free volume that are required for the interchange of particles of different species become rare quickly at high density while collective motion to relax density still remains possible. Therefore, if anything, we expect the slowing down of fractionation compared to density relaxation at high densities to be \textit{more} pronounced than in our lattice model. Hence the effects we observe should be stronger in more realistic, off-lattice models, getting closer to an ideal two-stage scenario for the phase-separation kinetics.

\begin{figure}[h]
	\centering
	\includegraphics[width=\columnwidth]{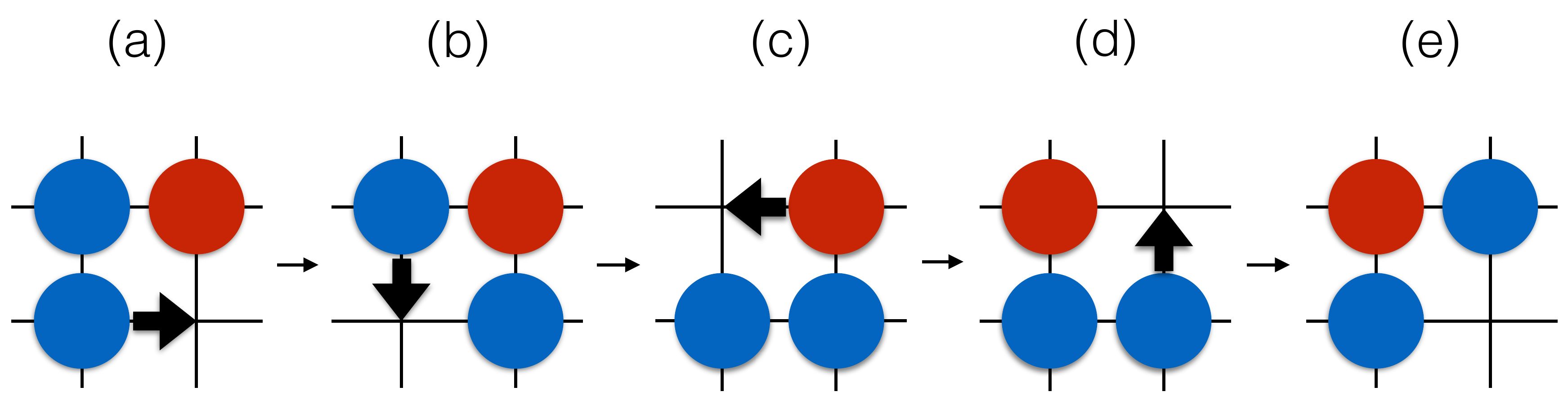}
	\caption{Sketch of vacancy-medidated interchange between two particles of different species.
In the first configuration a red particle is on the top-right site of the $2\times2$-lattice region, whereas a blue particle sits on the top-left site. In order for the subsystem to reach a configuration where one has instead a blue particle top-right and a red one top-left, four particle moves (particle-vacancy swaps)
are required, as shown by the sequence (a)--(e).
}
	\label{sketch}
\end{figure}

We illustrate how vacancies enable fractionation
in Fig.\ \ref{sketch}, which shows a sketch of a \textit{vacancy-mediated} interchange between two particles of different species.
A sequence of 4 particle moves within the small lattice portion shown ($2\times 2$ lattice sites) is sufficient to interchange the blue and red particle in the lattice row above the initial vacancy. (Note that in this sequence 
the two blue particles have also interchanged their positions, but as they are indistinguishable this is immaterial.)
Therefore, as fractionation requires interdiffusion of particles of different species, it will be able to proceed locally as long as vacancies are present. As a particle swap can be accomplished with a moderate number of elementary particle moves, fractionation cannot become arbitrarily slow compared to density relaxation--which locally requires a single particle move--as claimed above.
This is the likely reason why we do not see an `ideal' two-stage scenario, including e.g.\ the fact that at early times the spinodal dynamics within the quenched spinodal region already has a component of fractionation rather than consisting of the growth of pure density fluctuations.

Of course off-lattice models at the particle level are difficult to deal with theoretically, so before studying phase-separation dynamics one would aim to derive from them approximate representations in terms of a 
continuum field theory such as the so-called Model B or rather a polydisperse variant thereof.
\cite{doi:10.1080/00018739400101505,hohenberg1977theory}
For a monodisperse system and with stochastic fluctuations neglected, this would give an equation of the (generalized) Cahn\textendash Hilliard type\cite{cahn1958free} 
\begin{equation}
		\frac{\partial \rho\left(\mathbf{r},t\right)}{\partial t}=\nabla \, \cdot \left[\mathscr{M}\left(\rho\left(\mathbf{r},t\right)\right)\nabla \, \frac{\delta F\left[\rho\left(\mathbf{r},t\right)\right]}{\delta \rho\left(\mathbf{r},t\right)}\right]
		\label{CH}
	\end{equation}
where $F$ is the free energy functional, $\rho\left(\mathbf{r},t\right)$ is the continuous density field, and $\mathscr{M}\left(\rho\left(\mathbf{r},t\right)\right)$ is a density-dependent mobility; within a simple approximation scheme\cite{marconi1999dynamic} one finds $\mathscr{M}\left(\rho\left(\mathbf{r},t\right)\right)=\rho\left(\mathbf{r},t\right)$.

In the polydisperse case, the analogue of eqn (\ref{CH}) involves a mobility matrix with $M\times M$ entries. (In fact, our kinetic eqn (\ref{kineticequations}) can be cast in the form of a discrete Cahn\textendash Hilliard equation, generalized to polydisperse fluids and inhomogeneous mobilities; in the linear version, eqn (\ref{LinearDMFT}), the generalized Cahn\textendash Hilliard form is clear.)
Determining such an entire mobility matrix from an off-lattice model is a considerable challenge, also because the results might differ for size versus interaction polydispersity (similarly to the differences found in the mobility coefficients between systems with length and chemical polydispersity, in ref.\ \citenum{ignacio}). One route to deriving such Cahn-Hilliard-like equations would be 
dynamical density functional theory (DDFT); it should be possible to adapt this to a fully polydisperse scenario and then use it to test Warren's two-stage scenario.\cite{archer2009dynamical,archer2005dynamical}

A further theoretical challenge is the incorporation of stochastic effects, in order to be able to describe nucleation and growth dynamics. The simplest way of achieving this would be to add Langevin noise to either our lattice dynamics (\ref{kineticequations}) or a continuum description like
eqn (\ref{CH}).
This however requires a consistent, quantitative way of adding multiplicative Langevin noise to the deterministic equations. The need for a quantitative accurate stochastic description lies in the fact that one wants to \textit{compare} equilibration timescales between systems placed in different regions of the phase diagram, which means one has to assess the competition between nucleation and growth dynamics on the one hand and spinodal decomposition on the other.
For instance, for systems placed within R2 (see Table \ref{thetab}) one would like to compare the timescale for nucleation and growth in stage 1 (which should have a fast intrinsic time scale because it does not require fractionation, but could be slowed down by large nucleation barriers) to spinodal dynamics in stage 2 (intrinsically slow because it requires fractionation, but not slowed down by activation barriers).
A model being able to describe quantitatively both types of phase ordering dynamics would clearly be valuable here. This could potentially be obtained
from the lattice-based theory presented here via systematic coarse-graining. \cite{PhysRevLett.100.015702} Alternatively, fluctuating hydrodynamics\cite{jack2014geometrical,bertini2015macroscopic,dean1996langevin,kim2014equilibrium} is a possible avenue for deriving models incorporating stochasticity, though whether this can be implemented for polydisperse dynamics and would give a reasonable quantitative account of the physics remains an open question.

In addition to stochastic effects we have also neglected hydrodynamic interactions due to the solvent. This is commonly done throughout the literature on polydisperse dynamics, but in the case of dense systems, which are especially pertinent in the context of the two-stage phase-ordering scenario, they may play an important role.\cite{archer2009dynamical}
	
Our work in this paper could be extended to investigate the relaxation dynamics of polydisperse fluids in porous materials, generalising previous work done by Peter Monson and others. \cite{Monson2001, Monson2003, Monson2013} In fact, although the dynamical mean-field theory calculations developed in those papers are restricted to $M=1$ and $M=2$, they are very similar to the ones presented here. For application to porous materials, we would mainly need to adapt our approach to cover given pore geometries.
In this scenario, a natural question would be `what is the impact of slow fractionation on the relaxation dynamics in porous materials?'. This is potentially one of the simplest future research topics to pursue if one uses our current theory as a starting point.
	
Finally, we point out that one could use the framework developed here to investigate yet other problems, e.g.\ the dynamics of polydisperse wetting or the phase-separation behaviour of polydisperse systems of active particles. It would also be interesting to develop perturbative or scaling approaches for the dynamics of systems with small polydispersity, and to consider alternative ways of obtaining mean-field approximations, for example by explicitly taking the limit of high spatial dimension or considering a Kac-like setup with long-range interactions.\cite{activepolydispersity,peterwetting,evans,Kac}

\section{Acknowledgments}
PdC acknowledges financial support from CNPq, Conselho Nacional de Desenvolvimento Cient\'{i}fico e Tecnol\'{o}gico -- Brazil. PS acknowledges the stimulating research environment provided by the EPSRC Centre for Doctoral Training in Cross-Disciplinary Approaches to Non-Equilibrium Systems (CANES, EP/L015854/1).




\bibliography{Biblio} 
\bibliographystyle{rsc} 

\vspace{0.75cm}

\appendix
\section{Monotonic decrease of the free energy}
\label{monoticity}
Here we show that in the kinetic PLG model with both particle-vacancy and direct particle swaps (with Glauber-like rates, arbitrary $M$, arbitrary overall composition, and arbitrary attempt rates) the free energy obeys $dF/dt\leq0$. As we will see, the derivation is not as straightforward as in the case of a binary mixture without direct particle swapping, which was discussed in ref.\ \citenum{Plapp99}. The beginning is identical, though. As the free energy $F$ is a function of all dynamical variables $p^\alpha_i$ we can write
\begin{equation}
\frac{dF}{dt}=\sum_i\sum_\alpha\frac{\partial F}{\partial p^\alpha_i}\frac{dp^\alpha_i}{dt}
\end{equation}
which, combined with the definitions for the chemical potential and for the current, can be rewritten as
\begin{equation}
\frac{dF}{dt}=-\sum_\alpha\sum_i\sum_{j\in \partial i} \mu^\alpha_iJ_{ij}^\alpha
\ =\ -\sum_\alpha\sum_{\langle i,j\rangle} (\mu^\alpha_i J_{ij}^\alpha + \mu^\alpha_j J_{ji}^\alpha).
\end{equation}
Using the fact that $J_{ij}^\alpha=-J_{ji}^\alpha$ we can write
\begin{equation}
\frac{dF}{dt} = - \sum_{\alpha}\sum_{\langle i,j\rangle}  \left(\mu^\alpha_i - \mu^\alpha_j\right)J^{\alpha}_{ij}.
\end{equation}
Defining a two-species current $J^{\alpha\gamma}_{ij}$ in the obvious way using
\begin{equation}
J^{\alpha}_{ij}=\sum_{\gamma\geq0}J^{\alpha\gamma}_{ij}
\end{equation} 
transforms this further into
\begin{equation}
\frac{dF}{dt} = - \sum_{\langle i,j\rangle} \sum_{\alpha,\gamma} J^{\alpha\gamma}_{ij} \left(\mu^\alpha_i - \mu^\alpha_j\right).
\end{equation}
The sum can be extended to $\alpha\geq0$ because $\mu^0_i=0$, $\forall i$. Also note that $J^{\alpha\gamma}_{ij}$ is antisymmetric under interchange of $\alpha$ and $\gamma$. After adding the same expression with $\alpha$ and $\gamma$ swapped and using this antisymmetry, we have
\begin{equation}
\frac{dF}{dt} = - \frac{1}{2} \sum_{\langle i,j\rangle} \sum_{\alpha,\gamma} J^{\alpha\gamma}_{ij} 
\left[\left(\mu^\alpha_i - \mu^\alpha_j\right) - \left(\mu^\gamma_i - \mu^\gamma_j\right)\right]
\label{dfdt}
\end{equation}
However, the term in square brackets can be written as $\ln(C^\alpha_i C^\gamma_j) - \ln(C^\alpha_j C^\gamma_i)$, where we have defined the local chemical activity $C^\alpha_i$ as $C^\alpha_i = \exp(\mu^\alpha_i/T)$. This has the same sign as
$J^{\alpha\gamma}_{ij}$ because the two-species current can be written as a positive quantity multiplied by $C^\alpha_i C^\gamma_j - C^\alpha_j C^\gamma_i$. (This fact can be seen from the mean-field expression for the energy change, eqn (\ref{deltaHMF}),
which enters the rates via 
$\exp(\Delta H_{ij}^{\alpha\gamma}/T) = p_i^{\alpha}p_j^{\gamma}C_i^{\gamma}C_j^{\alpha}/( p_i^{\gamma}p_j^{\alpha}C_i^{\alpha}C_j^{\gamma}) $.) Therefore, each link $\langle i,j\rangle$ and each pair of $(\alpha,\gamma)$ makes a negative contribution to $dF/dt$. One can also phrase the last step in the other direction, i.e.\ by writing 
\begin{equation}
J^{\alpha\gamma}_{ij} = -\mathscr{M}^{\alpha\gamma}_{ij} \left[\left(\mu^\alpha_j - \mu^\alpha_i\right) - \left(\mu^\gamma_j - \mu^\gamma_i\right)\right]
\end{equation}
and checking that the site-dependent mobility $\mathscr{M}^{\alpha\gamma}_{ij}$ is positive. This makes sense because a particle swap is driven by how the difference in $\alpha$-chemical potential between sites $i$ and $j$ compares to the corresponding difference in $\gamma$-chemical potential. Note that while the factor $1/2$ in eqn (\ref{dfdt}) may looks somewhat unexpected, antisymmetry of the current $J^{\alpha\gamma}_{ij}$ again shows that $(\alpha,\gamma)$ and $(\gamma,\alpha)$ make the same contribution, so one could eliminate the $1/2$ again and restrict the summation to ordered pairs of species labels, say $\alpha<\gamma$.

\section{Derivation of growth rates} 
\label{growthratesappendix}
In this appendix we derive the linearized mean-field equations of motion and derive from them the expressions for the spinodal growth rates. Consider a homogeneous system of overall composition defined by a list of species densities $\left\lbrace p^{\alpha} \mid \alpha=1,\dots,M \right\rbrace$. The system is perturbed by small fluctuations of the densities:
\begin{equation}
p^{\alpha}_i=p^{\alpha}+\delta^{\alpha}_{i} \quad\forall \alpha
\end{equation}
where $\delta^{\alpha}_{i}\ll1$. In a linear expansion we can write eqn (\ref{deltaHMF}) as
\begin{equation}
\begin{aligned}
\bigl\langle\Delta H_{ij}^{\alpha\gamma}\bigr\rangle =  \sum\limits_{\beta}^{}\left(\sum\limits_{k\in \partial i}^{}\epsilon_{\alpha\beta}\delta_{k}^{\beta}-\sum\limits_{l\in \partial j}^{}\epsilon_{\alpha\beta}\delta_{l}^{\beta}\right)\\ - \sum\limits_{\beta}^{}\left(\sum\limits_{k\in \partial i}^{}\epsilon_{\gamma\beta}\delta_{k}^{\beta}-\sum\limits_{l\in \partial j}^{}\epsilon_{\gamma\beta}\delta_{l}^{\beta}\right).
\label{deltaHwithdeltas}
\end{aligned}
\end{equation}
Now remember our kinetic eqns (\ref{kineticequations})
\begin{equation*}
\begin{aligned}
\frac{dp_i^{\alpha}}{dt}=-\sum\limits_{j\in\partial i}^{}\sum\limits_{\gamma=0}^{M}\left[ \frac{p_i^{\alpha} p_j^{\gamma}	w^{\alpha\gamma}}{1+\exp{\left(\bigl\langle\Delta H_{ij}^{\alpha\gamma}\bigr\rangle/T\right)}}- \frac{p_j^{\alpha}p_i^{\gamma}	w^{\alpha\gamma}}{1+\exp{\left(\bigl\langle\Delta H_{ji}^{\alpha\gamma}\bigr\rangle/T\right)}}\right].
\end{aligned}
\end{equation*}
Using a simplified notation in which we drop the superscripts $\alpha$ and $\gamma$, the expression between the curly brackets can be written as
\begin{equation}
A_{ij}\frac{1}{1+\exp{\left(\Delta E_{ij}/T\right)}}-A_{ji}\frac{1}{1+\exp{\left(\Delta E_{ji}/T\right)}}
\label{twosigmoids}
\end{equation}
where $A_{ij}=w^{\alpha\gamma}p_i^{\alpha} p_j^{\gamma}$ and $\Delta E_{ij}=\bigl\langle\Delta H_{ij}^{\alpha\gamma}\bigr\rangle$.
As a consequence of linearization, the arguments in the exponentials are also small; hence, each sigmoid linearizes as $1/\left(1+\exp\left(x\right)\right) = \left(1/2\right)\left(1-x/2+\dots\right)$, and then we can write (\ref{twosigmoids}) in linearized version as
\begin{equation}
	\frac{1}{4}\left[2\left(A_{ij}-A_{ji}\right)-\frac{\Delta E_{ij}}{T}\left(A_{ij}+A_{ji}\right)\right]
\end{equation}
where we have used the fact that $\Delta E_{ij}=-\Delta E_{ji}$. Now eqn (\ref{kineticequations}) becomes
\begin{equation}
\begin{aligned}
\frac{d\delta_i^{\alpha}}{dt}=-\sum\limits_{j\in\partial i}^{}\sum\limits_{\gamma=0}^{M}\frac{w^{\alpha\gamma}}{4}\Biggl\lbrace 2 \left(p_i^{\alpha} p_j^{\gamma} - p_j^{\alpha}p_i^{\gamma}\right)\bigr.
\bigl.-\frac{\bigl\langle\Delta H_{ij}^{\alpha\gamma}\bigr\rangle}{T} \left(p_i^{\alpha} p_j^{\gamma} + p_j^{\alpha}p_i^{\gamma}\right)\Biggr\rbrace.
\end{aligned}
\label{15}
\end{equation}
As $\bigl\langle\Delta H_{ij}^{\alpha\gamma}\bigr\rangle$ is already linear in $\delta$'s, the last factor can be replaced by $2p^\alpha p^\gamma$. Linearizing also the first term then gives for
the time evolution of the fluctuations
\begin{equation}
\begin{aligned}
\frac{d\delta_i^{\alpha}}{dt}{}=
&-\sum\limits_{j\in\partial i}^{}\sum\limits_{\gamma=0}^{M}\frac{w^{\alpha\gamma}}{4}\Biggl\lbrace 2 \left[p^{\alpha}\left(\delta_j^{\gamma}-\delta_i^{\gamma}\right)\right.\biggr.\\
&\left.-p^{\gamma}\left(\delta_j^{\alpha}-\delta_i^{\alpha}\right)\right]
\biggl.-\frac{\bigl\langle\Delta H_{ij}^{\alpha\gamma}\bigr\rangle}{T} 2 p^{\alpha}p^{\gamma}\Biggr\rbrace.
\label{16}
\end{aligned}
\end{equation}
The first part can be rewritten as
\begin{equation}
\begin{aligned}
-\sum\limits_{j\in\partial i}^{}\sum\limits_{\gamma=0}^{M}\frac{w^{\alpha\gamma}}{4}\left\lbrace 2 \left[p^{\alpha}\left(\delta_j^{\gamma}-\delta_i^{\gamma}\right)-p^{\gamma}\left(\delta_j^{\alpha}-\delta_i^{\alpha}\right)\right]\right\rbrace\\
=-\Delta_{\rm d}\sum\limits_{\gamma=0}^{M}\frac{w^{\alpha\gamma}}{2T}p^{\alpha}p^{\gamma} \left(\frac{T}{p^{\gamma}}\delta_i^{\gamma}-\frac{T}{p^{\alpha}}\delta_i^{\alpha}\right)
\end{aligned}
\end{equation}
where $\Delta_{\rm d}$ is the discrete Laplacian, defined for any site-dependent quantity $g$ as $\Delta_{\rm d}g_i=\sum\limits_{j\in \partial i}^{}\left(g_{j}-g_i\right)$. For the second part, one can obtain
\begin{equation}
\begin{aligned}
-\sum\limits_{j\in\partial i}^{}\sum\limits_{\gamma=0}^{M}\frac{w^{\alpha\gamma}}{4}\left\lbrace-\frac{\bigl\langle\Delta H_{ij}^{\alpha\gamma}\bigr\rangle}{T} 2 p^{\alpha}p^{\gamma}\right\rbrace\\
=\Delta_{\rm d}\sum\limits_{\gamma=0}^{M}\frac{w^{\alpha\gamma}}{2T}p^{\alpha}p^{\gamma} \sum\limits_{\beta}^{}\left(\epsilon_{\gamma\beta}-\epsilon_{\alpha\beta}\right)\left(\Delta_{\rm d}+z\right)\delta_{i}^{\beta}
\end{aligned}
\end{equation}
where $z$ is the lattice coordination number and we used the fact that
\begin{equation}
	\begin{aligned}
		\sum\limits_{j\in\partial i}\left(\sum\limits_{k\in\partial i}\delta_k^{\beta}-\sum\limits_{l\in\partial j}\delta_l^{\beta}\right)
		=-\Delta_{\rm d}\left(\Delta_{\rm d}\delta_{i}^{\beta}+z\delta_{i}^{\beta}\right)
	\end{aligned}
\end{equation}
which can be easily derived in a few steps.
Bearing in mind that summations over $\beta$ do \textit{not} include $\beta=0$ and defining a homogeneous mobility $\mathscr{M}^{\alpha\gamma}=\frac{w^{\alpha\gamma}}{2T}p^{\alpha}p^{\gamma}$, we have altogether 
\begin{equation}
\begin{aligned}
\frac{d\delta_i^{\alpha}}{dt}={}
&\Delta_{\rm d}\sum\limits_{\gamma=0}^{M}\mathscr{M}^{\alpha\gamma}\sum\limits_{\beta}^{}\biggl\lbrace -\frac{T}{p^{\gamma}}\delta_{\gamma\beta}+\frac{T}{p^{0}}\delta_{\gamma 0}+\frac{T}{p^{\alpha}}\delta_{\alpha\beta}\biggl.\\
&\biggr.+\left(\epsilon_{\gamma\beta}-\epsilon_{\alpha\beta}\right)\left(\Delta_{\rm d}+z\right)\biggr\rbrace\delta_{i}^{\beta}.
\end{aligned}
\end{equation}
Note that the second term in the curly braces here comes from the identity
$\delta_i^0=-\sum_\beta \delta_i^\beta$, which is a consequence of the local hard core constraint (\ref{hardcore}).

Alternatively, using the local chemical potential (\ref{chempot}) linearized about the homogeneous state, 
\begin{equation}
\frac{d\delta_i^{\alpha}}{dt}=\sum\limits_{\gamma=0}^{}\mathscr{M}^{\alpha\gamma}\Delta_{\rm d}\Bigl(\mu_{i}^{\alpha}-\mu_{i}^{\gamma}\Bigr).
\label{mainequation}
\end{equation}

We can now proceed to solve the linearized dynamical equations of motion. As a homogeneous system is invariant under translation with respect to the lattice vectors, solutions of the linearized equations are superpositions of solutions of this form
\begin{equation}
\delta_{j}^{\alpha}=\delta p^{\alpha}\exp\left[i\mathbf{k}\cdot\mathbf{x}_j+\omega t\right]
\label{planewavesecondappearance}
\end{equation}
where it will be clear from the context that the letter $i$ here refers to the imaginary unit $i\equiv\sqrt{-1}$. Also, $\mathbf{k}$ is the fluctuation wave vector and $\mathbf{x}_j$ is the position in real space of a lattice site $j$. Moreover, $\omega$ is the spinodal decomposition growth rate and $\delta p^{\alpha}$ is the amplitude of the spinodal fluctuation associated with species $\alpha$. Inserting this form of the solution into our eqn (\ref{mainequation}) will give us an equation for $\omega$. In order to have a useful relation, however, the application of the discrete Laplacian operator $\Delta_{\rm d}$ to a $\delta_{j}^{\alpha}$ needs to be translated into some commutative operation so that we can deal with the $\exp\left[i\mathbf{k}\cdot\mathbf{x}_j+\omega t\right]$ properly, as follows:
\begin{equation}
\begin{aligned}
\Delta_{\rm d}\delta_{j}^{\alpha}{}
&=\delta p^{\alpha}\exp\left(\omega t\right)\sum\limits_{l\in \partial j}^{}\left[\exp\left(i\mathbf{k}\cdot\mathbf{x}_l\right)-\exp\left(i\mathbf{k}\cdot\mathbf{x}_j\right)\right]\\
&=\delta p^{\alpha}\exp\left(i\mathbf{k}\cdot\mathbf{x}_j+\omega t\right)\left[\sum\limits_{l\in \partial j}^{}\exp\left(i\mathbf{k}\cdot\mathbf{x}_l-i\mathbf{k}\cdot\mathbf{x}_j\right)-z\right]\\
&=\delta p^{\alpha}\exp\left(i\mathbf{k}\cdot\mathbf{x}_j+\omega t\right)\left[z+A-z\right]=A\delta_{j}^{\alpha}
\end{aligned}
\end{equation}
where (in the case of a $2$-dimensional system) we have $A=A(\mathbf{k})=-4\sin^2\left(k_{x}a/2\right)-4\sin^2\left(k_{y}a/2\right)$. Then this $A(\mathbf{k})$, or just $A$, replaces $\Delta_{\rm d}$, and eqn (\ref{mainequation}) can be written as an equation that no longer involves site-dependent quantities
\begin{equation} 
\begin{aligned}
\omega \delta p^{\alpha}={}
&\sum\limits_{\gamma=0}^{M}\mathscr{M}^{\alpha\gamma}\sum\limits_{\beta}^{}
\biggl\lbrace T\left(-\frac{\delta_{\gamma\beta}}{p^{\gamma}}+\frac{\delta_{\gamma0}}{p^{0}}+\frac{\delta_{\alpha\beta}}{p^{\alpha}}\right)\biggr.\\
&\biggl.+\left(\epsilon_{\gamma\beta}-\epsilon_{\alpha\beta}\right)\left(A+z\right)\biggr\rbrace A\delta p^{\beta}.
\end{aligned}
\label{omegaalpha}
\end{equation}
As we did in the main text, let us now set $w^{\alpha0}=w_0$ and $w^{\alpha\beta}=w_s$, for any $\alpha\neq0$ and $\beta\neq0$, where $w_0$ and $w_s$ are constant attempt rates associated with particle-vacancy and particle-particle exchanges, respectively. We next derive two equations involving $\omega$. The first is obtained by summing up eqn (\ref{omegaalpha}) over all species $\alpha$. This gives
\begin{equation}
	\omega \delta \rho=\frac{Aw_0}{2}\left[\delta\rho-\frac{A+z}{T}(1-\rho)\rho_1\delta \rho_1\right]
\end{equation}
where $\delta\rho\equiv\delta\rho_0$ and
\begin{equation}
	\rho_{n}=\sum\limits_{\alpha}^{}\sigma_\alpha^{n}  p^{\alpha}
\end{equation}
\begin{equation}
	\delta\rho_{n}=\sum\limits_{\alpha}^{}\sigma_\alpha^{n}\delta p^{\alpha}
\end{equation}
are the moment densities (and their fluctuations), and we have used $\epsilon_{\alpha\beta}=\sigma_\alpha\sigma_\beta$. For the case with a continuous polydisperse attribute, the summations are integrals. The second equation is obtained by multiplying eqn (\ref{omegaalpha}) by $\sigma_\alpha$ and then summing it over all species. The result is 
\begin{equation}
\begin{aligned}
\omega \delta \rho_1={}
&\frac{Aw_0}{2}\left[\rho_1\delta\rho+\left(1-\rho\right)\delta\rho_1-\frac{A+z}{T}\left(1-\rho\right)\rho_{2}\delta \rho_1\right]
\\
&+\frac{Aw_s}{2}\left[\rho\delta\rho_1-\rho_1\delta\rho+\frac{A+z}{T}\left(\rho_1^2-\rho_{2}\rho\right)\delta \rho_1\right].
\end{aligned}
\end{equation}
 By using that $\delta \rho$ and $\delta\rho_1$ should not both be nonzero in the spinodal dynamics and combining the two equations, one can now find $\omega(\mathbf{k})$. One obtains two non-trivial branches $\omega(\mathbf{k})$; we picked the largest one, which is eqn (\ref{growthrates}). There are a further $M-2$ eigenvalues (and eigenvectors) with $\delta\rho=\delta\rho_1=0$, leading to $M-2$ trivial branches such that all have identical negative eigenvalues. Explicitly, these are given by $\omega=\frac{A}{2}\left[ w_0(1-\rho)+w_s\rho\right]$ but as we are looking for the fastest growing modes, they can be ignored.
\end{document}